\documentclass[preprint,12pt,sort&compress]{elsarticle}
\usepackage[a4paper,margin=1in]{geometry}
\usepackage{setspace}
\usepackage{amsmath}
\usepackage{amssymb}
\usepackage{hyperref}
\usepackage{graphicx}
\usepackage{caption}
\usepackage{subcaption}
\usepackage{color}
\usepackage{algorithm}
\usepackage{algpseudocode}
\usepackage{pifont}
\usepackage{dblfloatfix} 
\usepackage{soul}
\usepackage{xcolor}

\soulregister\cite7
\soulregister\ref7
\soulregister\eqref7
\soulregister\pageref7
\sethlcolor{yellow}
\makeatletter
\def\SOUL@hlpreamble{%
   \setul{\dp\strutbox}{\dimexpr\ht\strutbox+\dp\strutbox\relax}%
   \let\SOUL@stcolor\SOUL@hlcolor
   \SOUL@stpreamble
}

\journal{ Materials \& Design }
\begin{document}

\begin{frontmatter}

\title{A novel physics-regularized interpretable machine learning model for grain growth  }
\tnotetext[t1]{{This manuscript has been co-authored by UT-Battelle, LLC, under contract DE-AC05-00OR22725 with the US Department of Energy (DOE). The publisher acknowledges the US government license to provide public access under the DOE Public Access Plan (http://energy.gov/downloads/doe-public-access-plan)}.}
\author[1]{Weishi Yan}
\author[1]{Joseph Melville}
\author[2]{Vishal Yadav}
\author[1]{Kristien Everett}
\author[2]{Lin Yang}
\author[3]{Michael S. Kesler}
\author[2]{Amanda R. Krause}
\author[2]{Michael R. Tonks}
\author[1]{Joel B. Harley \corref{cor1}}\ead{Joel.Harley@ufl.edu}
\cortext[cor1]{Corresponding author} 
\address[1]{Department of Electrical and Computer Engineering, University of Florida, Gainesville, FL}
\address[2]{ Department of Materials Science and Engineering, University of Florida, Gainesville, FL}
\address[3]{Oak Ridge National Laboratory, Oak Ridge, TN}
\begin{abstract}
Experimental grain growth observations often deviate from grain growth simulations, revealing that the governing rules for grain boundary motion are not fully understood. A novel deep learning model was developed to capture grain growth behavior from training data without making assumptions about the underlying physics. The Physics-Regularized Interpretable Machine Learning Microstructure Evolution (PRIMME) model consists of a multi-layer neural network that predicts the likelihood of a point changing to a neighboring grain. Here, we demonstrate PRIMME’s ability to replicate two-dimensional normal grain growth by training it with Monte Carlo Potts simulations. The trained PRIMME model’s grain growth predictions in several test cases show good agreement with analytical models, phase-field simulations, Monte Carlo Potts simulations, and results from the literature. Additionally, PRIMME’s adaptability to investigate irregular grain growth behavior is shown. Important aspects of PRIMME like interpretability, regularization, extrapolation, and overfitting are also discussed.
\end{abstract}

\end{frontmatter}

\section{Introduction}\label{sec:introduction}

Microstructure  significantly influences the physical properties of metallic and ceramic materials \cite{dillon2016importance}. Many efforts have been made to develop mathematical descriptions of the spatial and temporal evolution of polycrystalline microstructures at elevated temperatures used in processing or application. During grain growth in isotropic systems, the velocity of a migrating grain boundary $v_b$ is defined by a driving force $F_b$ and a grain boundary mobility $M_b$ \cite{humphreys1997unified}:
\begin{equation}
    v_b = M_b F_b. \label{v_GB}
\end{equation} 
The primary driving force is the grain boundary energy $\gamma_b$, such that $F_b = \gamma_b/R_b$, where $R_b$ is the radius of curvature of the boundary. The simplest case of grain growth is the shrinking of a circular grain embedded in a matrix for which the rate of change of the grain area $A$ is defined as
\begin{equation}
    \frac{\partial A}{\partial t} = 2 \pi r \frac{\partial r}{\partial t} = - 2 \pi M_b \gamma_b,
    \label{eq:dAdt_circle}
\end{equation}
where $r$ is the radius of the grain and $\partial r/ \partial t = - v_b$ from Eq.~\eqref{v_GB}. This expression can be integrated with time to define the grain area as a function of time $t$:
\begin{equation}
    A(t) = A_0 - 2 \pi M_b \gamma_b t.
    \label{eq:A(t)}
\end{equation}
Following Burke and Turnbull analysis \cite{Burke1952}, evolution of the mean grain size $ \langle R \rangle$ of a polycrystalline grain structure is given as
\begin{equation}
     \langle R \rangle^2 - \langle R_o \rangle^2 = K t, \label{eq:av_gr_size_w_time}
\end{equation}
where $ \langle R_o \rangle$ is the initial mean grain size (the spherical equivalent radius of an arbitrarily shaped grain) and $K$ is the kinetic coefficient.  Another important aspect of  two-dimensional polycrystalline grain structure is the relationship between the number of sides of a grain $F$ and its area $A$, which is also known as the von Neumann-Mullins relation \cite{mullins1956two} and is given as
\begin{equation}
    \frac{dA}{dt} = -\frac{\pi}{3}  M_b \gamma_b \left(6-F\right).\label{eq:vonNeumannMullins}
\end{equation}
Accordingly, grains with more than six sides will grow and those with less than six will shrink, while grains with six sides will remain stable. 


Computational modeling is another approach to model polycrystalline grain growth and it has been carried out for nearly forty years. The earliest model of grain growth used the Monte Carlo Potts (MCP) method \cite{wu1982potts} to represent the local changes throughout a grain structure \cite{anderson1984computer,srolovitz1984computer}.  In this stochastic approach, the grain structure is divided into discrete grain sites and the grain assigned to each site randomly changes to neighboring grains based on a probability that is a function of the grain boundary energy. In addition, various deterministic methods have been developed to model grain growth.  In phase field grain growth methods, first used by Fan and Chen \cite{fan1997computer}, the grain structure is represented by continuous variable fields that have constant values within grains and smoothly transition between values at grain boundaries. The variables evolve with time to minimize a functional that defines the overall free energy of the system \cite{kim2014phase,miyoshi2017ultra,Miyoshi2021M,Chadwick2021,Moelans2022}. Cellular automata \cite{liu1996simulation,he2006computer,ding2006cellular,Xiong2021,Baumard2021},  front tracking \cite{frost1988two,lazar2010more,lazar2011more}, and level set methods \cite{elsey2009diffusion,Fausty2021} have also been used to model grain growth.

Analytical models assume that the material properties are isotropic. In reality, however, the kinetics of atomic movement depends on the anisotropic grain boundary energy ($\gamma_b$) and mobility ($M_b$) \cite{rollett1989simulation}. Similarly, computational models for grain growth are developed using assumptions and approximations. These assumptions and approximations result in errors when compared against experimental data \cite{mckenna2014grain}. A recent experimental finding has challenged the most accepted grain growth theories by revealing that there is no observed relationship between grain boundary velocity and curvature in polycrystalline Ni  \cite{bhattacharya2021grain}. In addition, simulations have to be solved numerically and can be computationally expensive for large numbers of grains. Therefore, an alternative and efficient computational approach that accurately mimics  grain growth experiments is needed.

Recently, machine learning modeling approaches have been tremendously successful in many scientific computational tasks by implementing statistical inference on a very large {scale  \cite{Carou_undated-go,Datta2021-un}.} In particular, deep learning has proven to be a powerful tool for analyzing dynamic systems \cite{qian2020lift}, including complex material microstructures \cite{bostanabad2018computational}. However, most of this work has focused on microstructure recognition \cite{chowdhury2016image} and reconstruction \cite{bostanabad2016stochastic}. There are limited studies that take advantage of state-of-the-art machine learning methods for modeling the dynamic evolution of microstructures. In prior work, de Oca Zapiain et al.\ implemented a surrogate model to facilitate phase field model predictions \cite{de2021accelerating}. However, in their approach, the application of machine learning is heavily dependent on the high-fidelity phase field model. Yang et al.\ applied convolutional recurrent neural networks to predict microstructural evolution of various complexities \cite{yang2021self}, but the black-box machine learning model simply mimics the simulator without providing meaningful insights about the underlying physics. Moreover, machine learning models trained on large, clean simulation data alone cannot be used for real-world experimental prediction, and it is extremely expensive to collect sufficient experimental microstructure data for proper training. Hence, a generalized machine learning framework that can be trained with simulations, guided by physics, and remain flexible enough to be tuned with new experimental knowledge is of high priority. 

The objective of the current work is to train a  deep neural network model to predict two-dimensional isotropic grain growth {(assuming all grain boundaries have the same energy and mobility)} and validate its results with analytical models of normal grain growth. This is a first step in a long-term vision: to develop an interpretable, physics-guided deep learning model for predicting spatio-temporal grain boundary migration in two and three dimensions. The remainder of this paper is structured as follows: First, the Physics-Regularized Interpretable Machine Learning Microstructure Evolution (PRIMME) architecture is presented. This PRIMME model is trained with simulated data (MCP simulations using the Stochastic Parallel PARticle Kinetic Simulator (SPPARKS)). PRIMME results are then compared with analytical  models and MCP and  { phase field (PF) } simulations. Finally, various aspects of PRIMME are discussed in detail.  In future work, this two-dimensional isotropic PRIMME model, which now contains interpretability and integrated physics, will be extended to analyze complex three-dimensional experimental grain growth with anisotropic grain boundary energy and mobility.

\section{Model Development}
\label{sec:model}

Designing a robust machine learning architecture for grain growth is not immediately straightforward. Ideally, the machine learning will directly predict the next  grain structure using the previous grain structure. The input would be an array containing the grain number at each site and the output would be another array of grain numbers, representing the evolved grain structure after some period of time. However, this goal does not cleanly fit into the traditional classification or regression machine learning archetypes \cite{james2013introduction}. As a result, a new architecture is necessary for grain growth. 

Our new architecture is inspired by the dynamic processes in the Monte Carlo Potts model and concepts from deep reinforcement learning \cite{mnih2013playing}. Rather than learn grain numbers directly, it learns how they act given their local grain neighborhood. We assume that each site has a finite number of actions at each time step. An action is defined as a site adopting the grain number of a neighbor (i.e., the site ``flips''). This new architecture first learns the most likely action for each site and then applies this action, which may or may not change the assigned grain. 

Furthermore, rather than learn the action directly, this new architecture learns the likelihood for each action (similar to a regression problem). It then chooses the action with the maximal likelihood (similar to a classification problem). This approach provides flexibility because particular actions can be constrained based on known physics or knowledge. It also increases interpretability, since the outcome is a mapping of the most likely actions for each site. To extend this architecture for more complex microstructure in the future, these two properties are highly desirable.

\begin{figure*}
\centering
 \includegraphics[width=\textwidth]{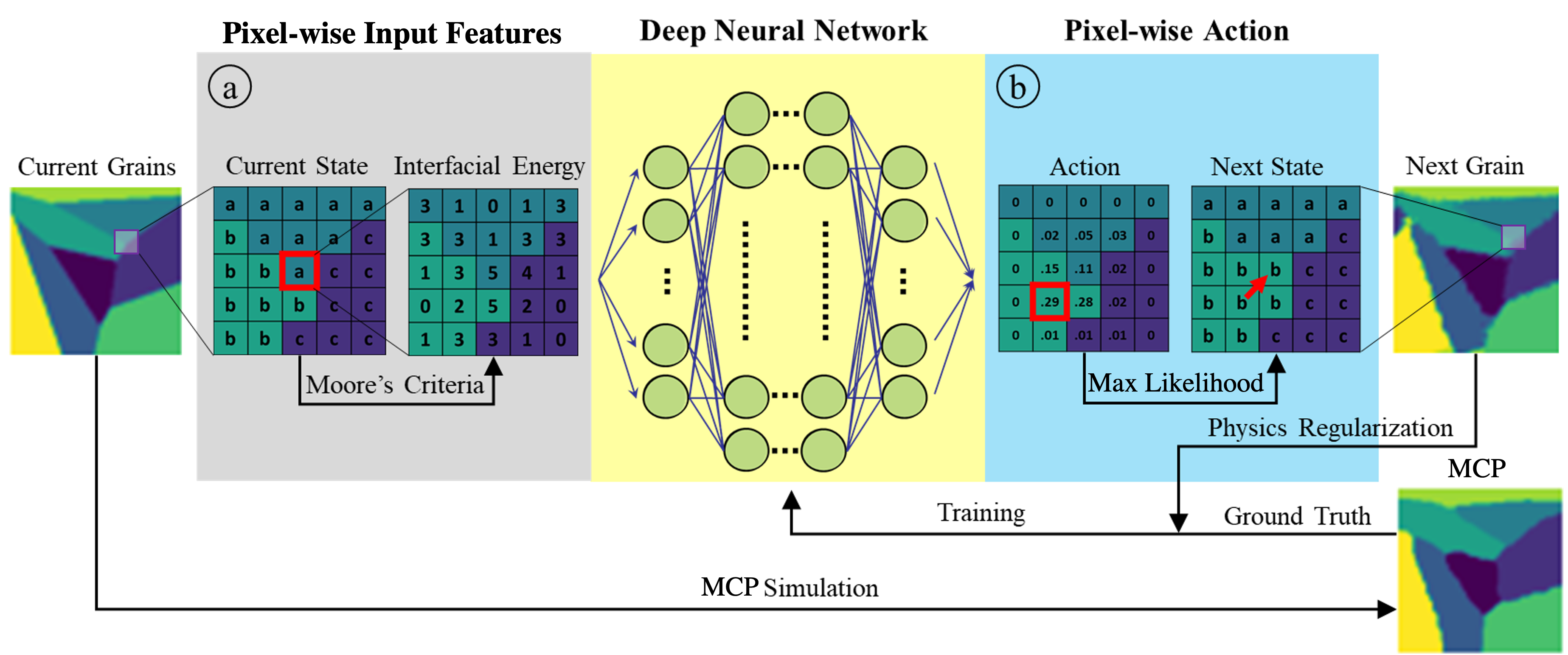}
 \caption{Summary of the PRIMME deep learning approach for grain growth simulation. (a). Boundary pixels are selected to create pixel-wise features. Kronecker delta of Moore’s neighbors are used as the deep neural network input. (b). The neural network outputs the action likelihood for a state around the center site. The maximum action likelihood is selected to flip the center site to corresponding grain number.}
 \label{FIG:flowchart}
\end{figure*}

Figure~\ref{FIG:flowchart} illustrates the complete framework of the proposed new architecture. In the framework, the microstructure is first divided into local neighborhoods and  physically relevant data features are extracted from each neighborhood. These features are taken as input into a neural network, which outputs the likelihood of taking each action. Subsequently,  the action with maximum likelihood is applied to obtain the evolved grain structure. A fully connected deep neural network is employed to learn actions due to its advantage in learning  complex, non-linear relationships within data. The following subsections detail each of these steps.

\subsection{Training Data Generation from a Monte Carlo Potts Model}
\label{ssec:sppark}

A MCP algorithm was used to model grain growth using the SPPARKS code \cite{cardona2009crossing}, assuming isotropic grain boundary properties. Each SPPARKS simulation was initialized with a $257 \times 257$ pixelated domain with 256 initial grains generated using a Voronoi diagram to assign the grain number for each site $s_i$. The initial grain structure was then evolved using the MCP method, simulating grain growth. Within one simulation step, each site in the image can evolve to one of its allowable neighbors based on a probability linked to the energy state given by the Hamiltonian
\begin{equation}
    \mathcal{H} = \sum_{i=1}^N \left[ \frac{1}{2} \sum_{j \in \mathcal{N}_i} \overline{\gamma}_b (1-\delta(s_i, s_j)) \right],\label{eq:SPPARKS_hamiltonian}
\end{equation}
where $N$ is total number of sites (i.e., pixels) in the image, $\mathcal{N}_i$ is the Moore's neighborhood of sites surrounding site $i$, $\overline{\gamma}_b$ is the isotropic average grain boundary energy, and $\delta(s_i, s_j)$ is the Kronecker delta
\begin{equation}
	\delta(s_i, s_j) = \begin{cases}
	1, & s_i = s_j \\
	0, & s_i \neq s_j \\
	\end{cases}. \label{eq:delta}
\end{equation}
Hence, the term $1-\delta(s_i, s_j)$ is $0$ when two sites are from the same grain and $1$ when they are from different grains.

The Hamiltonian represents the total interfacial energy contributed by the grain boundaries. The {probability $P$ of} flipping to an allowable neighbor {that results in an energy change $\Delta E$, if we assume all boundaries have the same mobility, is defined as}
\begin{equation}
	P(\Delta E) = \begin{cases}
	1, & \Delta E \leq 0 \\
	e^{\frac{-\Delta E}{kT}}, & \Delta E > 0 \\
	\end{cases},
\end{equation}
{where the computational temperature $kT$ is a hyperparameter that defines the intrinsic randomness of the Monte Carlo model \cite{anderson1984computer}}. In the training data, $kT=0.5$ was found {to roughen the boundaries enough} to avoid lattice pinning and {still result in a general decrease in the global energy to }provide normal growth behavior. We assume that the energy of all grain boundaries is equal with value $\overline{\gamma}_b=1$. Therefore, the SPPARKS simulations only consider local curvature and hence simulate normal curvature-driven parabolic grain growth.

\subsection{Neural Network Inputs and Outputs}
\label{ssec:action}

The input into the neural network is derived from images of grain number and is based on a modified Moore's neighbors criteria, where the observation can be extended to more than one immediate neighbor. Specifically at site $s_i$, the number of neighbors with a different grain number is calculated as
\begin{align}
    \mathcal{I}(s_i) = \sum_{j\in \mathcal{N}_n} 1-\delta(s_i, s_j),
    \label{eq:number_diff_neighbors}
\end{align}
where $\mathcal{N}_n$ is the set containing indices for the $n \times n$ neighborhood of pixels around site $s_i$. For this paper, we use $n=7$ for the network input.  Hence, if the original microstructure is a $257 \times 257$ pixelated image of grain numbers, then $\mathcal{I}_n(s_i)$ represents an image of the same dimensions (with spatial locations indexed by $i$), where each value is equal to the number of neighboring sites with a different grain number (Fig.~\ref{FIG:flowchart}a), resembling the Hamiltonian in Eq.~\eqref{eq:SPPARKS_hamiltonian}. The new image $\mathcal{I}(s_i)$ is then divided into overlapping $\mathcal{O} \times \mathcal{O}$ patches $\mathcal{P}(s_i, s_j)$ around site $s_i$. The index $s_j$ corresponds to the $\mathcal{O} \times \mathcal{O}$ neighboring sites. If the patch is at the edge of the image, we assume periodic boundary conditions. Each patch is vectorized into a  $\mathcal{O}^2 \times 1$ vector to become a single input to the neural network. The number of neural network inputs from one image is equal to the number of pixels in the image, $N$. When training, we randomize the order of patches but we train with a single step of the entire simulated microstructure before progressing to the next simulation. 

The output of the neural network is defined by the predicted action likelihoods around each site $s_i$. Therefore, the output is an $\mathcal{A}^2 \times 1$ vector that can form a $\mathcal{A} \times \mathcal{A}$ image  $Y(s_i, s_j)$ of a neighborhood around $s_i$. The index $s_j$ corresponds to the $\mathcal{A} \times \mathcal{A}$ neighboring sites. Each value defines the action likelihood that site $s_i$ will flip to the grain number associated with index $s_j$ (Figure~\ref{FIG:flowchart}b). We flip $s_i$ according to the highest action likelihood. After simultaneously applying an action to each site (rather than in a sequence), the grain structure is considered ``evolved'' to its next state.

\subsection{Neural Network Loss Function and Regularization}
\label{ssec:loss}

A custom loss function is used to robustly link the inputs to the action likelihood outputs. The loss for site $s_i^{(t)}$ at time step $t$ is then defined by a squared error term
\begin{equation}
    \label{eq:loss}
    L\left(s_i^{(t)} \right) = \frac{1}{|\mathcal{N}_{\mathcal{A}}|} \sum_{j \in \mathcal{N}_{\mathcal{A}}} \Bigg| Y\left(s_i^{(t)}, s_j^{(t)}\right) - \sum_{\tau=t}^{t+{N_t}} \left(\frac{1}{2} \right)^{\tau} \left[ \delta\left( s_i^{(\tau+1)},s_j^{(t)} \right) +  \lambda \, \Gamma\left( s_i^{(\tau)},s_j^{(\tau)} \right) \right] \Bigg|^2
\end{equation}
where $\mathcal{N}_{\mathcal{A}}$ is set of neighbors around $s_i^{(t)}$ in the action space, $|\mathcal{N}_{\mathcal{A}}|$ is the size of the set, $Y(s_i^{(t)}, s_j^{(t)})$ is the neural network's predicted action likelihood, and $\Gamma(s_i^{(t)},s_j^{(t)})$ is the physics-guided regularization with weight $\lambda$. {$\tau$ is the index used in the summation over the number of future time steps.} Inspired by the cumulative future rewards used in deep reinforcement learning \cite{mnih2015human}, we consider {a number of }future time steps {$N_t$} in our learning framework. If {$N_t=1$}, then the loss is only computed for the immediate next step. As {$N_t$} increases, we consider more future time steps.  In this paper, we use {$N_t = 4$} and $\lambda=1$. 

The first term within this sum $\delta( s_i^{(\tau+1)},s_j^{(t)}) $ is the direct action label. It trains the network to choose the true future grain number, according to the MCP model. The second term is a physics-guided regularization mechanism that encourages actions that decrease the number of neighbors with different grain numbers $\mathcal{I}(s_i)$ at each pixel site of the system. This regularization is defined by
\begin{align}
    \label{eq:regularization_label}
    \Gamma(s_i^{(t)}, s_j^{(t)}) &= 
    \frac{1}{8} \left[ \sum_{k \in \mathcal{N}_3}
     \delta \left( s_j^{(t)}, s_k^{(t+1)} \right) -
     \delta \left( s_i^{(t)}, s_k^{(t)} \right) \right] \; 
\end{align}
where $\mathcal{N}_3$ represents a $3 \times 3$ neighborhood. The first term in $\Gamma(s_i^{(t)}, s_j^{(t)})$ is the number of neighbors that match each of the possible new grain numbers $s_j^{(t)}$ in the next time step $s_k^{(t+1)}$. Hence, this value is high at site $s_j^{(t)}$ when more of the sites in $s_k^{(t+1)}$ are the same grain number. The second term gives the initial number of sites that match the center site before taking actions. The maximum value of our regularization is $1$, which occurs when site $s_i$ matches none of its neighbors (second term is $0$) and every neighbor has the same grain number (first term is $8$). The minimum value of our regularization is $-1$ when all sites in the neighborhood have the same grain number. Therefore, the regularization strongly discourages sites from flipping when most of its neighbors already have the same grain number.

{Note that the regularization only influences how the neural network is trained. It does not place hard constraints on the output nor is it directly computed as part of the neural network's output in testing. As a result, given sufficient and consistent training data, the neural network could learn irregular grain growth behavior where, for example, a small grain surrounded by larger ones could grow.}

\subsection{Neural Network Design}

The neural network architecture is shown in Table~\ref{tbl: model architecture}. Three hidden layers with Rectified Linear Unit (ReLU) activation are fully connected to take the $\mathcal{O}^2$ inputs from observation space. Batch normalization layers are added before the activation layers to improve learning speed and stability. A 25\% random dropout is performed to prevent the co-adaptation of neurons \cite{hinton2012improving}. The output layer contains $\mathcal{A}^2$ neurons with a sigmoid transformation to represent the likelihood of flipping actions within the action space. The input and output size, $\mathcal{O}^2$ and $\mathcal{A}^2$, respectively, are tunable. We chose an observation space and action space of $\mathcal{O} = \mathcal{A} = 17$ due to its balance of speed and performance.

\begin{table}[t!]
\centering
\caption{Neural Network Architecture}
\begin{tabular}{ p{2cm} p{4cm} p{1.3cm} }
 Layer  &Layer description &Activation\\
 \hline
 Flatten & Input Size = $\mathcal{O}^2$ & -\\[1mm]
 BatchNorm & Batch Normalization & - \\[1mm]
 Dense & Size = 1764 & ReLU\\[1mm]
 Dropout &  Drop Rate = 0.25 & -\\[1mm]
 BatchNorm & Batch Normalization & - \\[1mm]
 Dense & Size = 882 & ReLU\\[1mm]
 Dropout &  Drop Rate = 0.25 & -\\[1mm]
 BatchNorm & Batch Normalization & - \\[1mm]
 Dense & Size = 441 & ReLU\\[1mm]
 BatchNorm & Batch Normalization & - \\[1mm]
  Dense&  Output Size = $\mathcal{A}^2$  & Sigmoid \\[1mm]
 \hline
\end{tabular}

\label{tbl: model architecture}
\end{table}

\subsection{Training the Neural Network}

\begin{figure}[b!]
\centering
 		\includegraphics[width=0.5\textwidth]{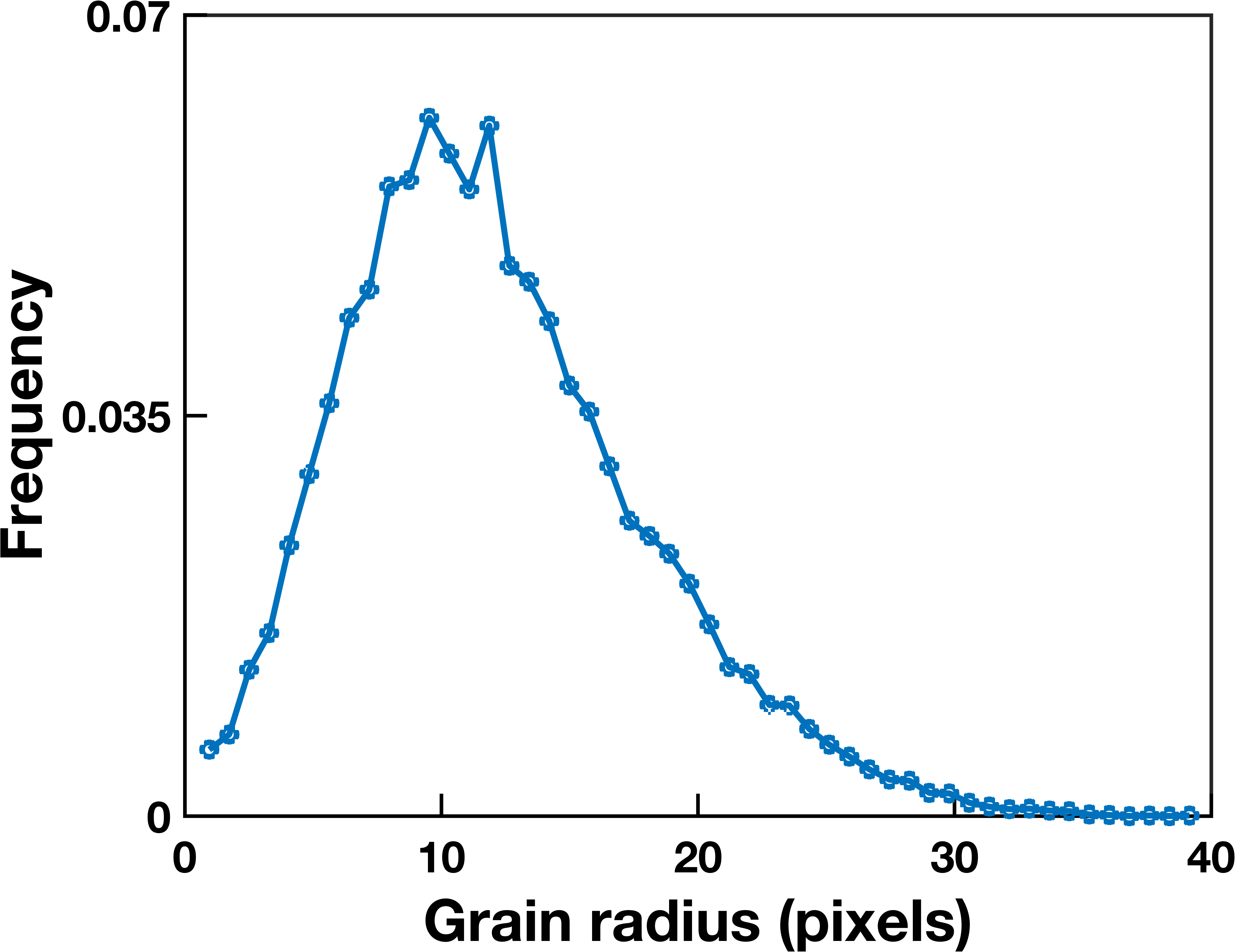}
   \caption{Training data grain sizes. Histogram of the grain sizes in the MCP training data.}   \label{FIG:training_size_distribution}

\end{figure}
The neural network was trained on results from the MCP simulations evolved for up to a maximum of 100 Monte Carlo steps (MCS). A total of $200$ simulations were run for training, each with unique initial conditions (resulting in $66,049$ neural network inputs per simulation, one for each site). For each simulation, we only considered one point of time $t$, to prevent overfitting to a single initial condition. Generating the MCP training data and training PRIMME required approximately 5 hours. Figure~\ref{FIG:training_size_distribution} illustrates the distribution of grain sizes in the MCP training data. It is skewed to the left and follows a log-normal distribution. Note that the training data only contains grains with radii ranging from 0 to 40 pixels.  This enables us to observe how well PRIMME can generalize the grain growth behavior to grains with sizes larger than those contained in the training data.

After the training is complete, PRIMME is ready to predict isotropic grain growth for any two-dimensional domain and for any initial grain structure. It can use both zero-flux or periodic boundary conditions. Like the MCP model,  PRIMME  is non-dimensional. However, unlike the MCP model, PRIMME is deterministic not stochastic. {Thus, the assumptions of the model are that all grain boundaries have the same energy and mobility (grain boundary migration is only driven by curvature), the domain is two-dimensional, and that the behavior is non-dimensional in space and time.}

\section{Results}
\label{sec:result_discussion}

In this section, a systematic microstructural analysis is carried out with the trained two-dimensional isotropic PRIMME model. First, we analyze the simplest case - evolution of a circular grain. Next, we qualitatively compare polycrystalline microstructure evolution from PRIMME to the MCP and phase field models. Finally, we investigate geometric and topological properties \cite{Barmak2013} of large-scale grain growth. We also compare PRIMME simulation results with analytical models for grain growth. The MCP simulations are carried out using SPPARKS. The phase field simulations are carried out using the Multiphysics Object-Oriented Simulation Environment (MOOSE) \cite{permann2020moose}, a parallel finite element framework developed by Idaho National Laboratory with a computationally efficient implementation \cite{permann2016order} of the grain growth model from Moelans et al. \cite{moelans2008quantitative}. The parameters for the PF simulation are as follows:  element size =$1\ \mu$m, a grain boundary mobility $M_b = 3.24\times10^{-11}$ m$^4$/(Js), energy $\gamma_b = 0.74$ J/m$^2$, interface width $=6\ \mu$m and time-step $=0.1$ sec.

\subsection{Evolution of a circular grain}
\begin{figure}[t!]
\centering
	\begin{subfigure}[b]{0.45\textwidth}
 		\includegraphics[width=\textwidth]{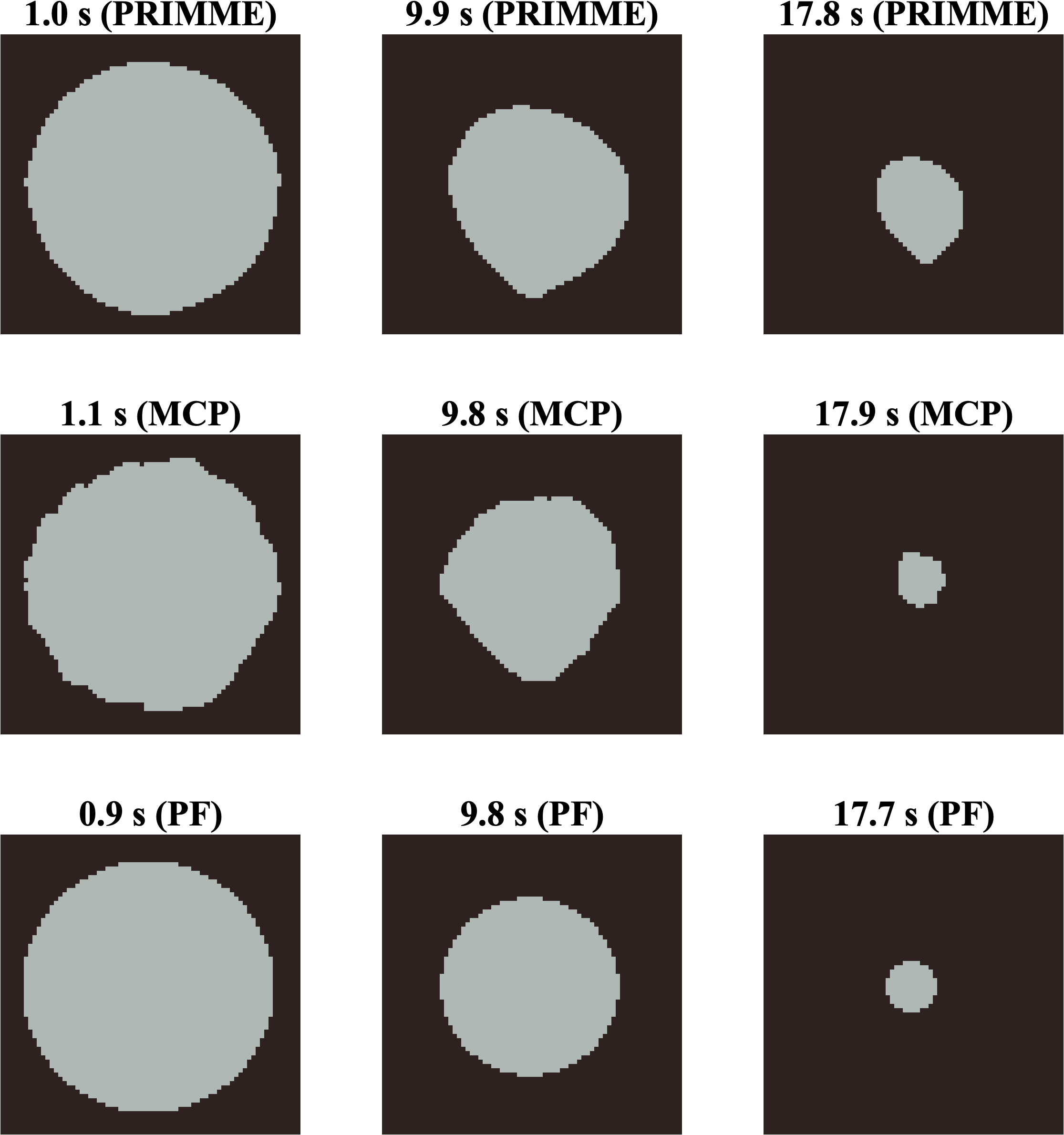} 		
	    \caption{\label{FIG:vanish_images}}
	\end{subfigure}	
	
	\begin{subfigure}[b]{0.45\textwidth}
 		\includegraphics[width=\textwidth]{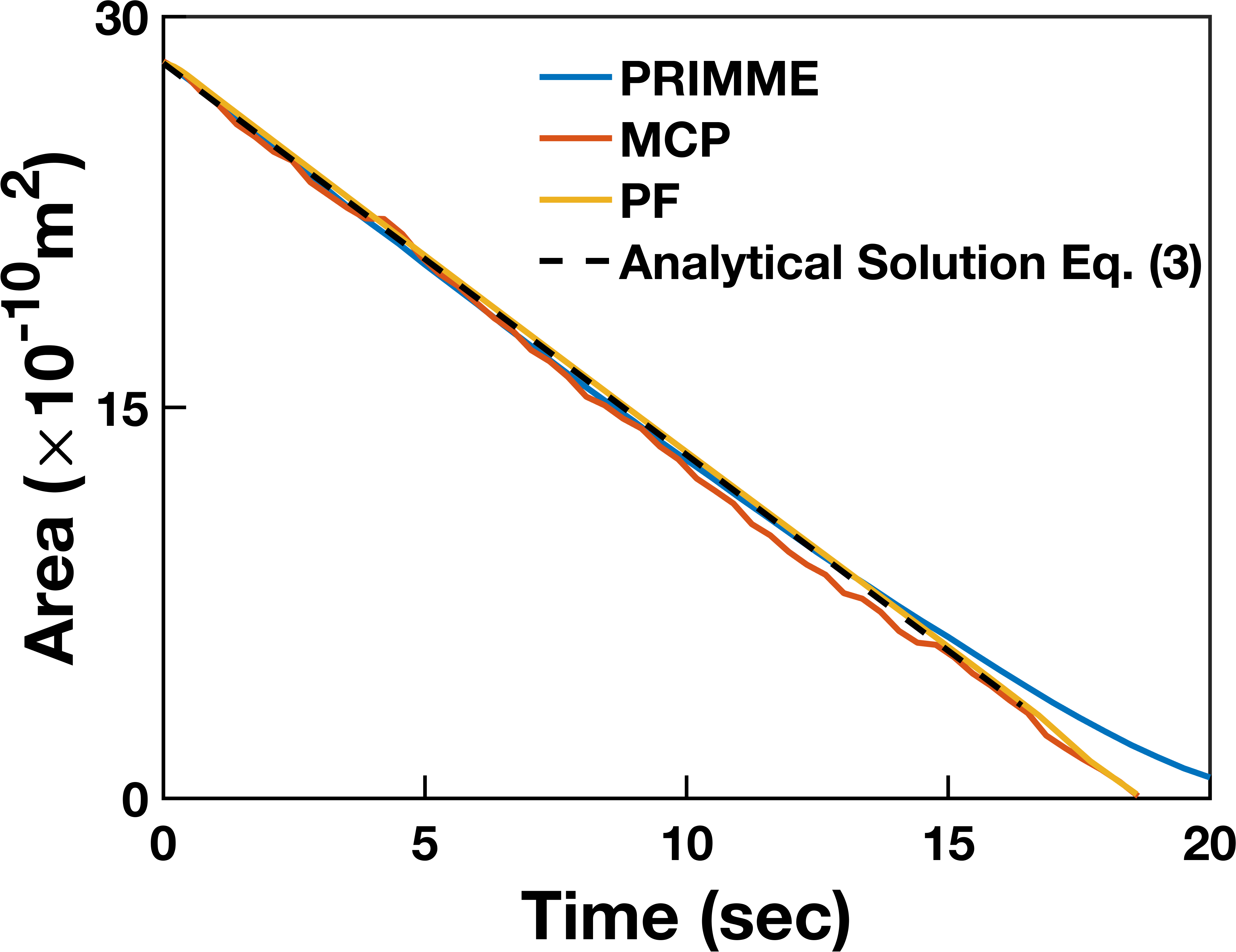} 		
	    \caption{\label{FIG:vanish_plot}}
	\end{subfigure}
	
   \caption{Evolution of a 30 $\mu$m radius circular grain in a $256\times256$ pixel matrix.   (a) Images of the shrinking grain from PRIMME, MCP and PF simulations. The time in PRIMME and MCP is scaled to real time by fitting to the analytical solution from Eq.~\eqref{eq:A(t)}.(b) The change in area of the circular grain with time from PRIMME, MCP, PF. Eq.~\eqref{eq:A(t)} is also plotted for reference.}
  \label{FIG:vanish}

\end{figure}

The evolution of circular grains embedded in a matrix simulated by PRIMME, MCP and PF models are shown in Fig.~\ref{FIG:vanish_images}. The initial radius of the circular grain for the PRIMME and MCP models is 30 pixels and the size of the matrix is $256 \times 256$ pixels. Similarly, the initial radius of the circular grain in the PF model is  30 $\mu$m and the size of the matrix is  $256\ \mu\mathrm{m}  \times 256\ \mu$m.  PRIMME is nondimensional in time and space, similar to the MCP model. However, since it evolves the grain structure differently than is done in the MCP implementation in SPPARKS, its evolution in time will be different. In order to compare all three simulations,  PRIMME and MCP results  are scaled using the analytical solution from Eq.~\eqref{eq:A(t)}. The scaling is determined by assuming 1 pixel = 1 $\mu$m  for PRIMME and MCP and the grain boundary mobility and energy are the same as for the PF model.   By fitting the circular grain results to Eq.~\eqref{eq:A(t)}, one PRIMME step is equal to $0.50$~s and on{e}  MC step is equal to $0.35$~s. This scaling procedure, where PRIMME and MCP results are scaled by fitting to analytical models or PF simulations, will be used in the following examples. 

The scaled grain area versus time plot for the PRIMME, MCP and PF models is shown in Fig.~\ref{FIG:vanish_plot}, and the analytical solution from Eq.~\eqref{eq:A(t)} is included for reference. All three models follow a linear relationship between area and time{, consistent with the analytical model and other isotropic grain growth models from the literature \cite{anderson1984computer,fan1997computer,liu1996simulation,frost1988two,elsey2009diffusion}}. PRIMME and MCP do not maintain a circular shape of the grain with time; the shape fluctuates in the MCP result but maintains a consistent oblong shape in the PRIMME result. {This behavior disagrees with other deterministic models from the literature that maintain a circular shape like the phase field result shown here \cite{fan1997computer,liu1996simulation,frost1988two,elsey2009diffusion}}. When the circle becomes very small, PRIMME results deviate from the analytical solution {due to the oblong shape of the grain. It takes longer to disappear than a circular grain}. Note that the PRIMME model was never trained on this initial condition but the shrinking behavior was reasonably described (though the shape was not), demonstrating that PRIMME can generalize beyond a Voronoi initial condition.

\subsection{Evolution of polycrystalline microstructure}

\begin{figure*}[b!]
\centering
 \includegraphics[width=\textwidth]{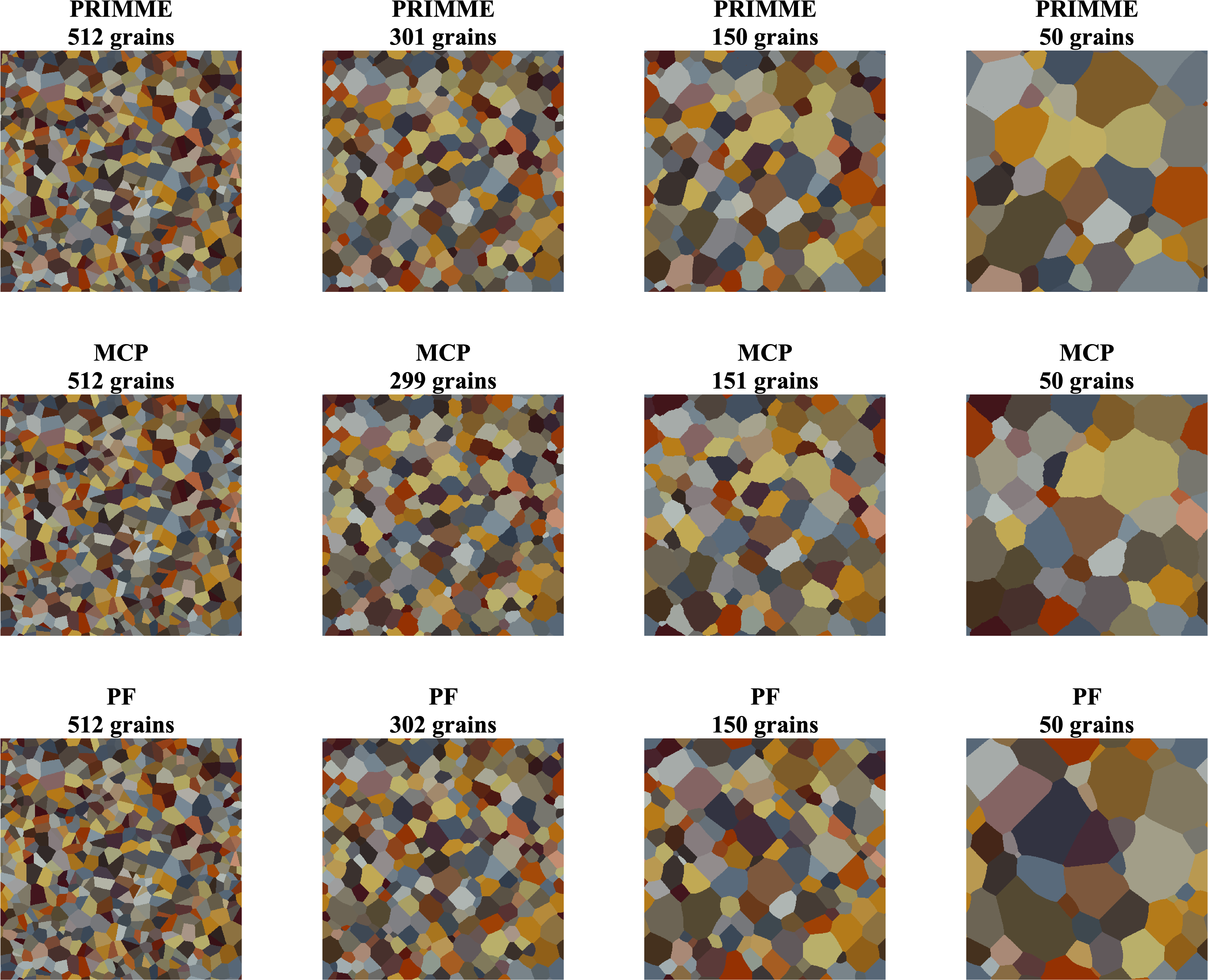}
 \caption{Comparison of the grain structure evolution predicted by PRIMME, MCP and PF in a $512\times512$ pixel domain with 512 initial grains. Grain structures with similar numbers of grains are compared from the three methods.}
 \label{FIG:growth}
\end{figure*}
Next, we compare the PRIMME simulation results for a polycrystalline grain structure with those from the MCP and PF models. We use a $512 \times 512$ pixel domain with $512$~initial grains generated by a Voronoi diagram, which has both more grains and a larger domain than used in the training set.  The initial condition is identical for the three methods. Other simulation details (like grain boundary energy, grain boundary mobility, etc.)  are identical to the circular grain case for all three models. PRIMME simulation is performed for $1000$ steps, which is larger than the $100$~maximum~steps used in the training data.

As shown in Fig.~\ref{FIG:growth}, the PRIMME simulation has grain structures typical for normal grain growth, with triple junctions converging towards three stable $120^{\circ}$ angles. PRIMME and MCP have similar grain structures when the number of grains $\approx300$. This is after 60 MCP steps, which is still within the range of the 100 steps used in the training data. The PF structure is somewhat less similar, consistent with previous comparisons between MCP and PF grain growth model comparisons \cite{tikare1998comparison,suwa2005computer}. As grain growth continues, the local grain structures predicted by the three methods diverge, though they still have similar microstructural characteristics. This behavior is expected as PRIMME is now extrapolating outside of the training data and MCP is stochastic. {The grain growth behavior predicted by PRIMME is in agreement with other isotropic grain growth models from the literature \cite{anderson1984computer,fan1997computer,liu1996simulation,frost1988two,elsey2009diffusion}.}

\subsection{Microstructural analysis of large-scale grain structure}
To quantify the performance of PRIMME in a case with sufficient number of grains to allow for an accurate statistical comparison with MCP and PF models, we simulate a $2400 \times 2400$ pixel ($\mu$m) domain with 20,000 initial grains generated using a Voronoi diagram. The grain growth is simulated using the PF method for $300$~s, which is equivalent to 1000 MCP steps. Thus, these large scale simulations have a domain size, number of grains, and number of steps that is far outside of the training data. We carry out a geometric analysis, comparing the change in the mean grain size and the grain size distribution, and a topological analysis, comparing the mean number of sides and the distribution of the number of sides. We compare the PRIMME results with our MCP and PF simulations and to previous simulation results.

However, we first compare the computational cost of the three simulations. It is difficult to directly compare the computational cost of PRIMME, MCP using SPPARKS, and PF using MOOSE since they use different approaches, different hardware, and different levels of parallelization. However, since both SPPARKS and MOOSE are widely-utilized and highly optimized engines for grain growth simulation, a rough comparison of wall clock time is valuable. Table~\ref{tbl: performance} shows the number and type of processor used for the simulations and the total wall time required. SPPARKS and MOOSE both run on CPUS, while PRIMME is GPU-enabled. The total wall time required for PRIMME is shorter than the times for SPPARKS and MOOSE: its time using two graphics cards is 63\% of the wall time required for SPPARKS using four 64-core processors and is 12\% of the wall time required for MOOSE using 40 32-core processors. 

\begin{table*}[b!]
\centering
\caption{Computational cost of PRIMME, SPPARKS and MOOSE for the 20,000 grain simulation.}
\begin{tabular}{ p{3.5cm} p{3.5cm} p{3.5cm} p{3.5cm} }
 Model  &PRIMME &SPPARKS &MOOSE\\
 \hline
 Processors & \textbf{2} Nvidia Quadro RTX 8000 & \textbf{4} AMD EPYC 7702 64-Core & \textbf{40} AMD EPYC 75F3 32-Core \\[1mm]
 Total wall time (s) & 4,839 & 7,776 & 40,590 \\[1mm]
 \hline
\end{tabular}

\label{tbl: performance}
\end{table*}
\subsubsection{Geometric analysis}
\begin{figure}[b!]
\centering
    \begin{subfigure}[b]{0.4\textwidth}
    \includegraphics[width=\textwidth]{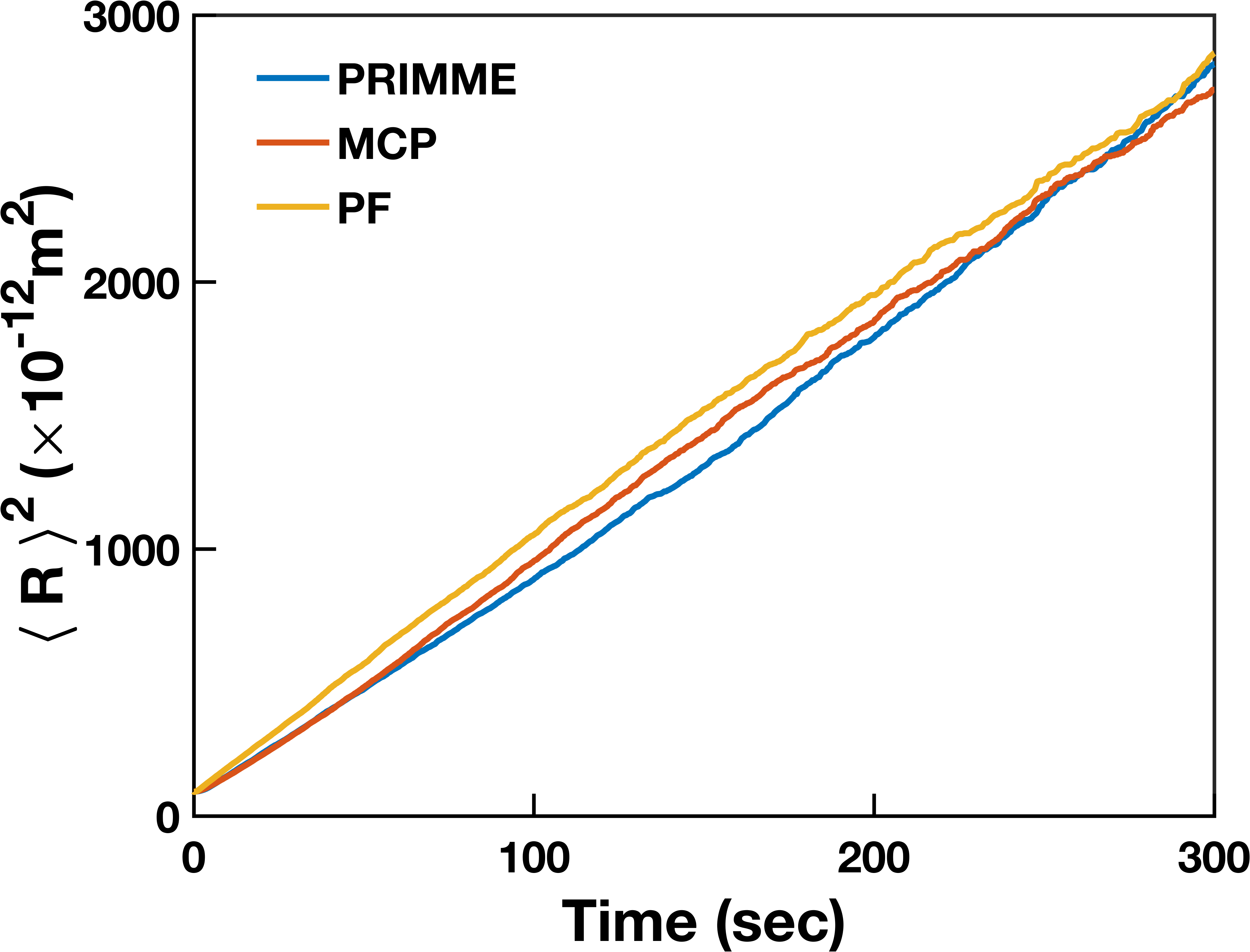}
    \caption{\label{fig:av_gr_area_vs_time}}
    \end{subfigure}
        \begin{subfigure}[b]{0.4\textwidth}
    \includegraphics[width=\textwidth]{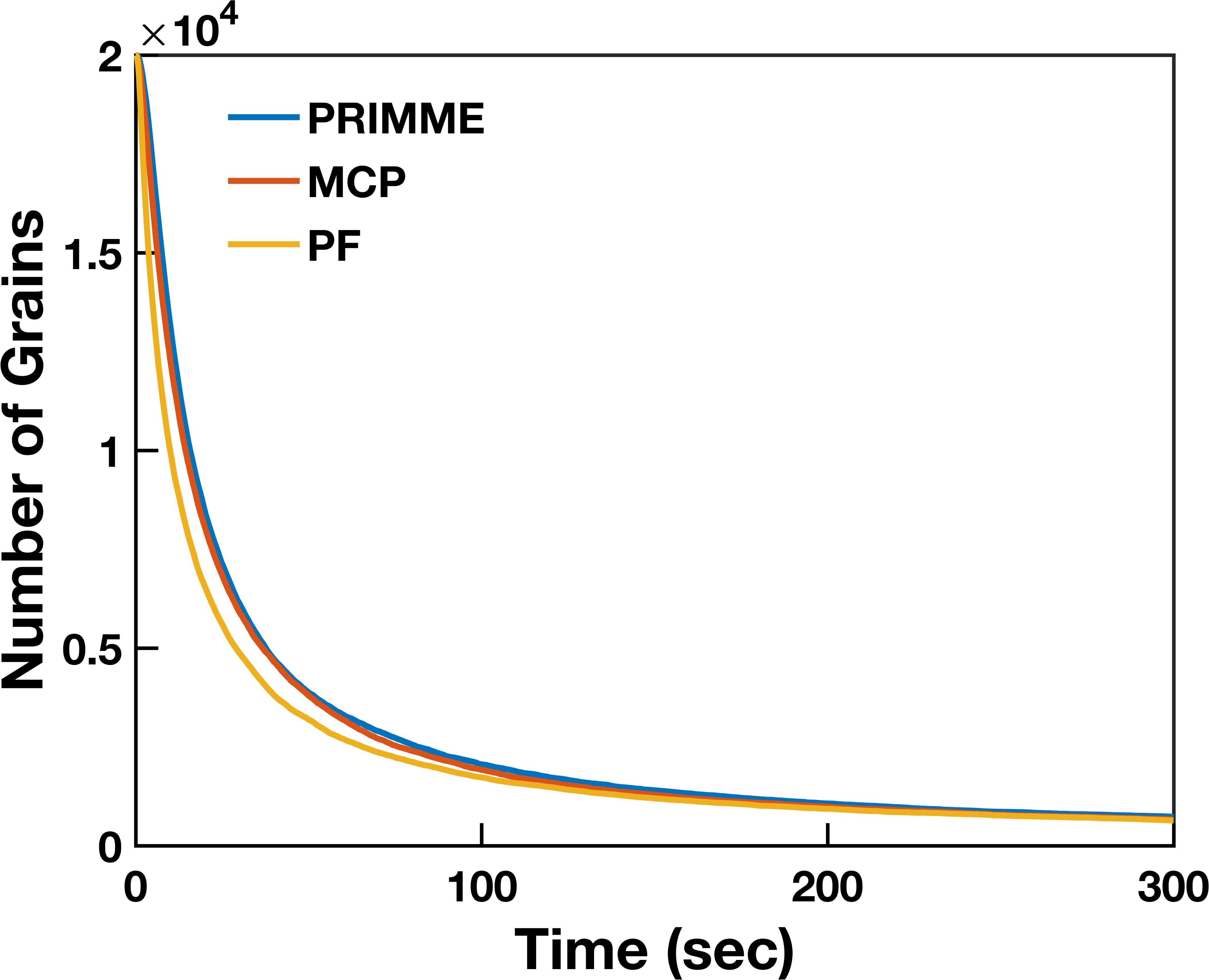}
    \caption{\label{fig:numg_rain_vs_time}}
    \end{subfigure}
    \begin{subfigure}[b]{\textwidth}
 		\includegraphics[width=0.32\textwidth]{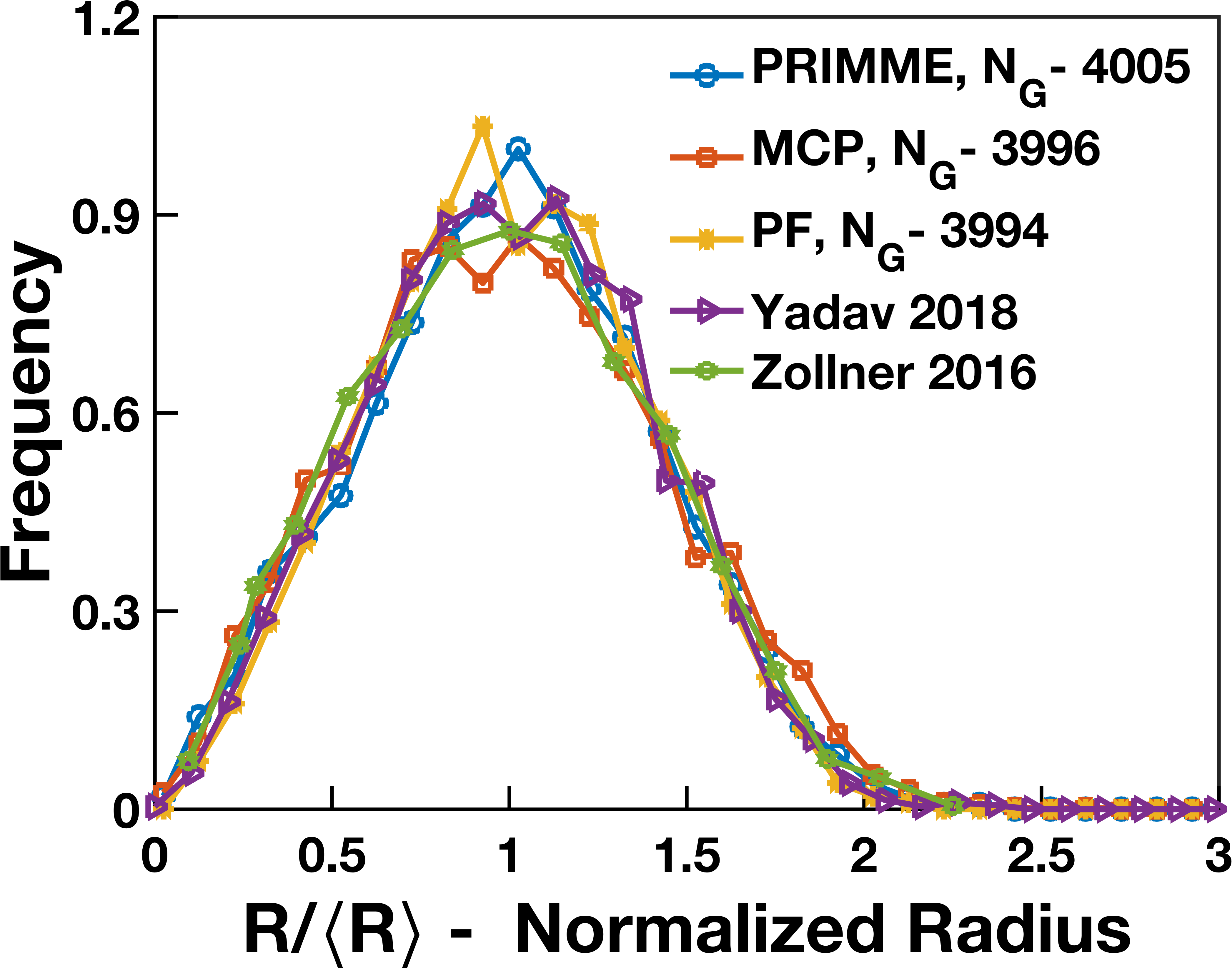} 
	    \includegraphics[width=0.32\textwidth]{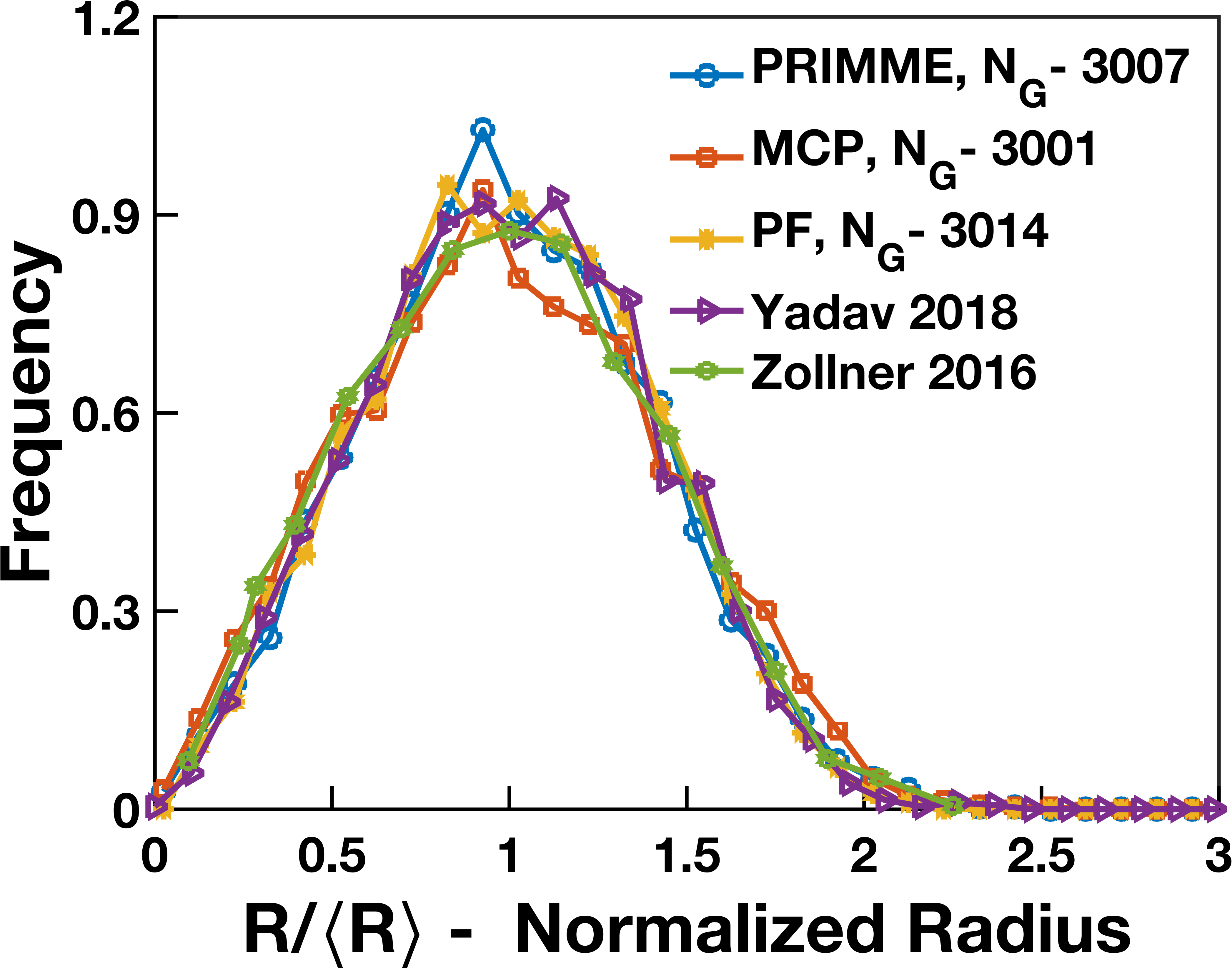} 	\includegraphics[width=0.32\textwidth]{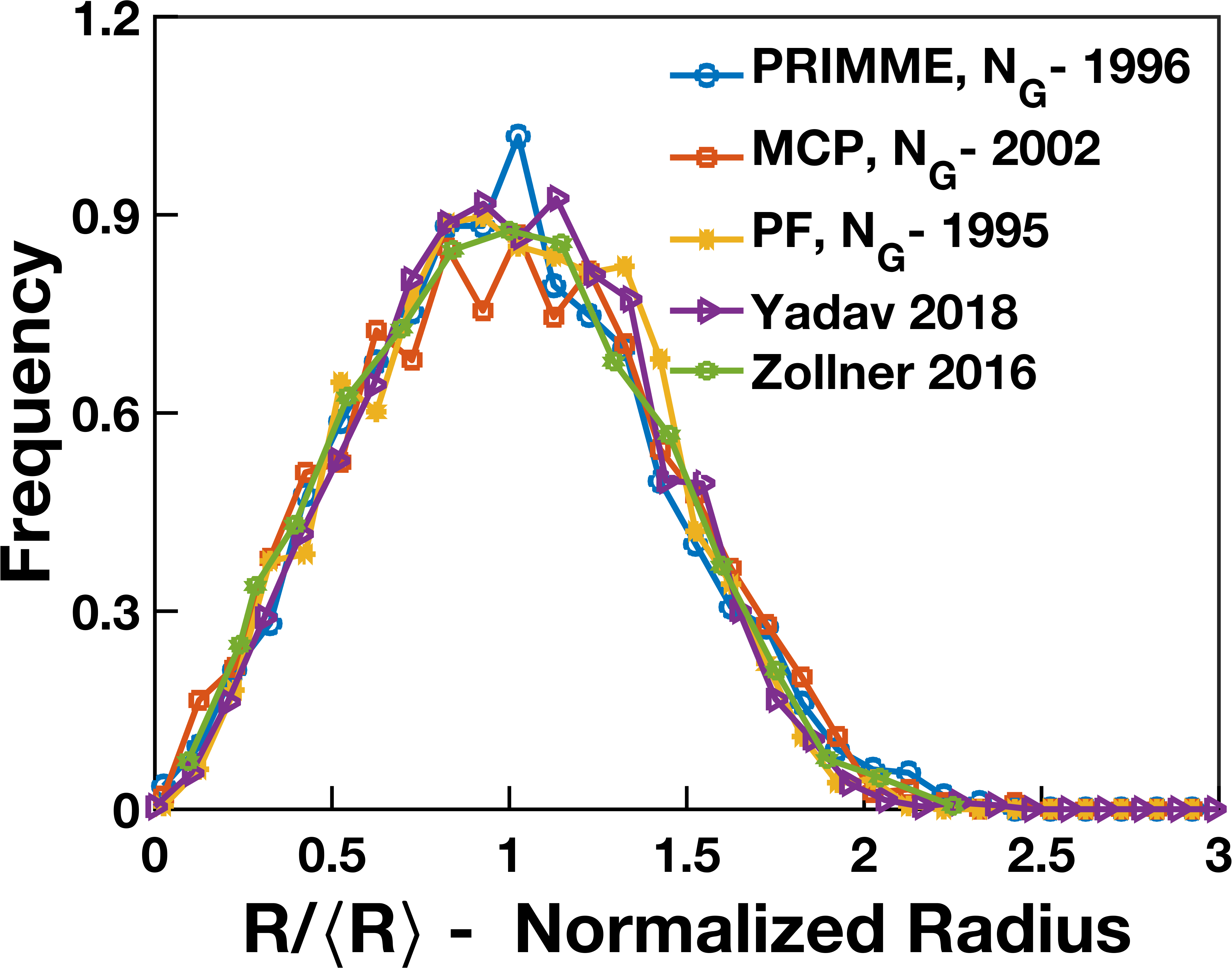} 	\caption{\label{fig:gr_size_distr}}
    \end{subfigure}
    \caption{Geometric analysis of the $2400\times2400$ pixel domain with 20,000 initial grains. (a) Evolution of the square of the mean grain size with time from PRIMME, MCP, and PF. The PRIMME and MCP results were scaled to real time by fitting their slope to the slope of the PF results. {(b) Evolution of number of grains with time.} (c) Comparison of the grain size distribution for approximately 4000, 3000, and 2000 grains from PRIMME, MCP, and PF. The results from Yadav 2018 \cite{Yadav2018} and Zollner 2016 \cite{Zollner2016} are also included, for reference.}
\end{figure}

The square of the average grain size with time from the PRIMME, MCP, and PF simulations is shown in Fig.~\ref{fig:av_gr_area_vs_time}. The PRIMME and MCP results were scaled to real time by fitting their slopes to the slope of the PF results. PRIMME accurately predicts the linear relationship between grain size and time (Eq.~\eqref{eq:av_gr_size_w_time}) and is very similar to that predicted by MCP and PF. {The evolution of number of grains  is shown in Fig.~\ref{fig:numg_rain_vs_time}.}

The distribution of the grain sizes for grain structures with approximately 4000, 3000, and 2000 grains from PRIMME, MCP, and PF are shown in Fig.~\ref{fig:gr_size_distr}. In normal isotropic grain growth, the distribution of the grain sizes normalized by the average grain size is self similar, meaning that it is constant over time. The grain size distribution predicted by PRIMME does not significantly change as the grain structure evolves. The shape of the grain size distribution predicted by PRIMME is very similar to that predicted by MCP and PF. In addition, it is {in good agreement with} the grain size distribution from large-scale two-dimensional MCP simulations \cite{Zollner2016} and PF simulations \cite{Yadav2018} from the literature.

\subsubsection{Topological analysis}
\begin{figure}[b!]
\centering
    \begin{subfigure}[b]{0.4\textwidth}
    \includegraphics[width=\textwidth]{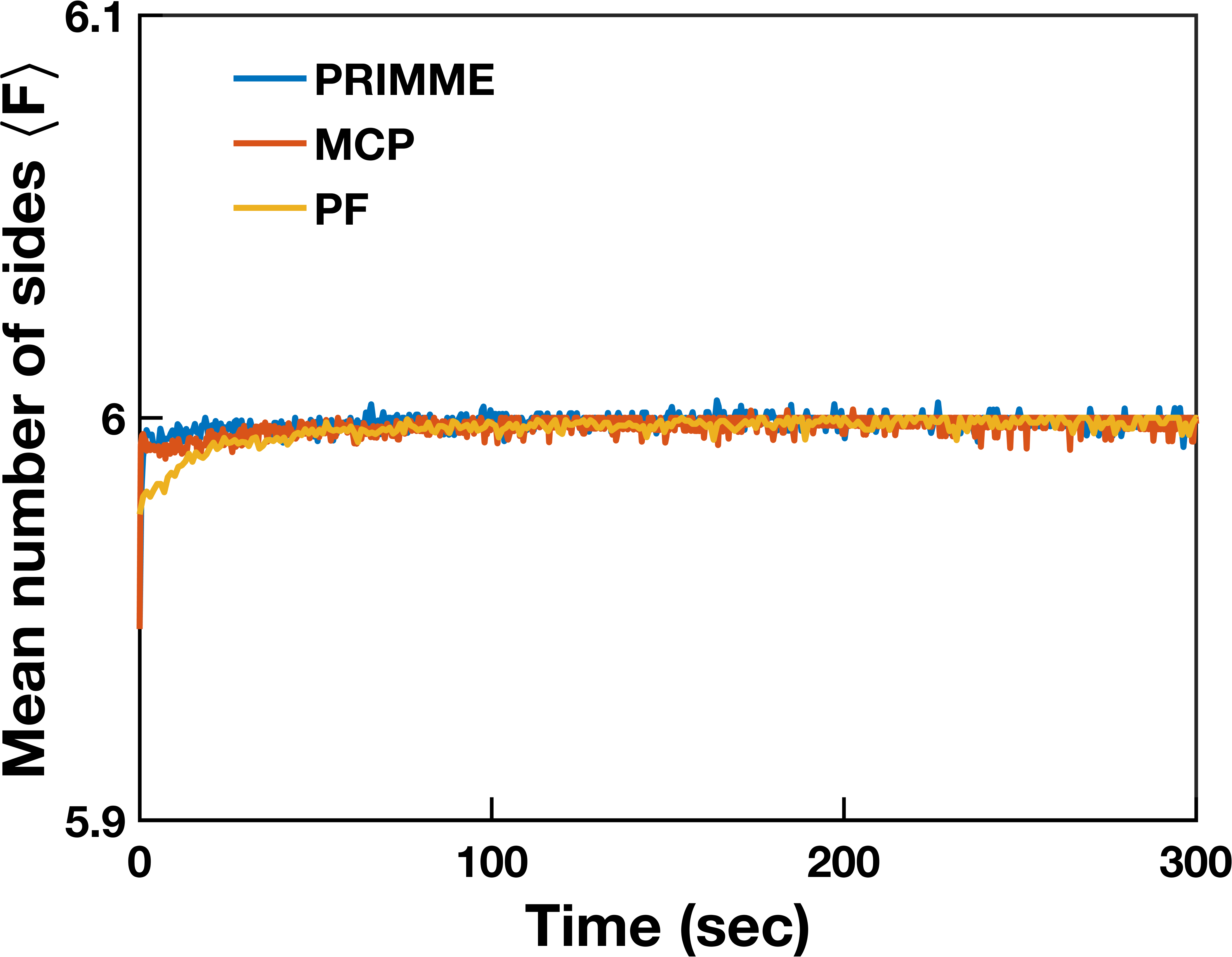}
    \caption{\label{fig:av_num_sides_vs_time}}
    \end{subfigure}
    \begin{subfigure}[b]{\textwidth}
 		\includegraphics[width=0.32\textwidth]{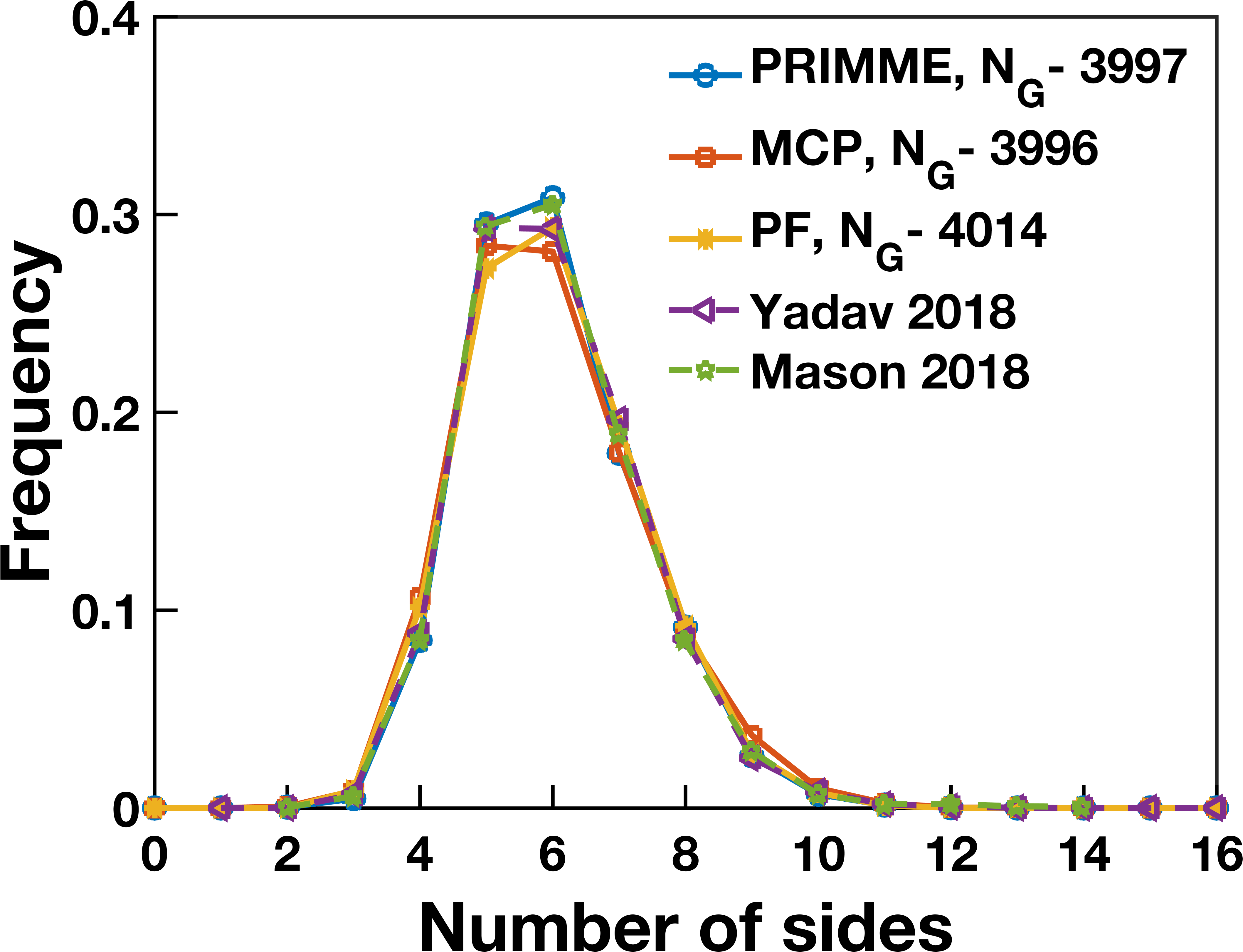} 
	    \includegraphics[width=0.32\textwidth]{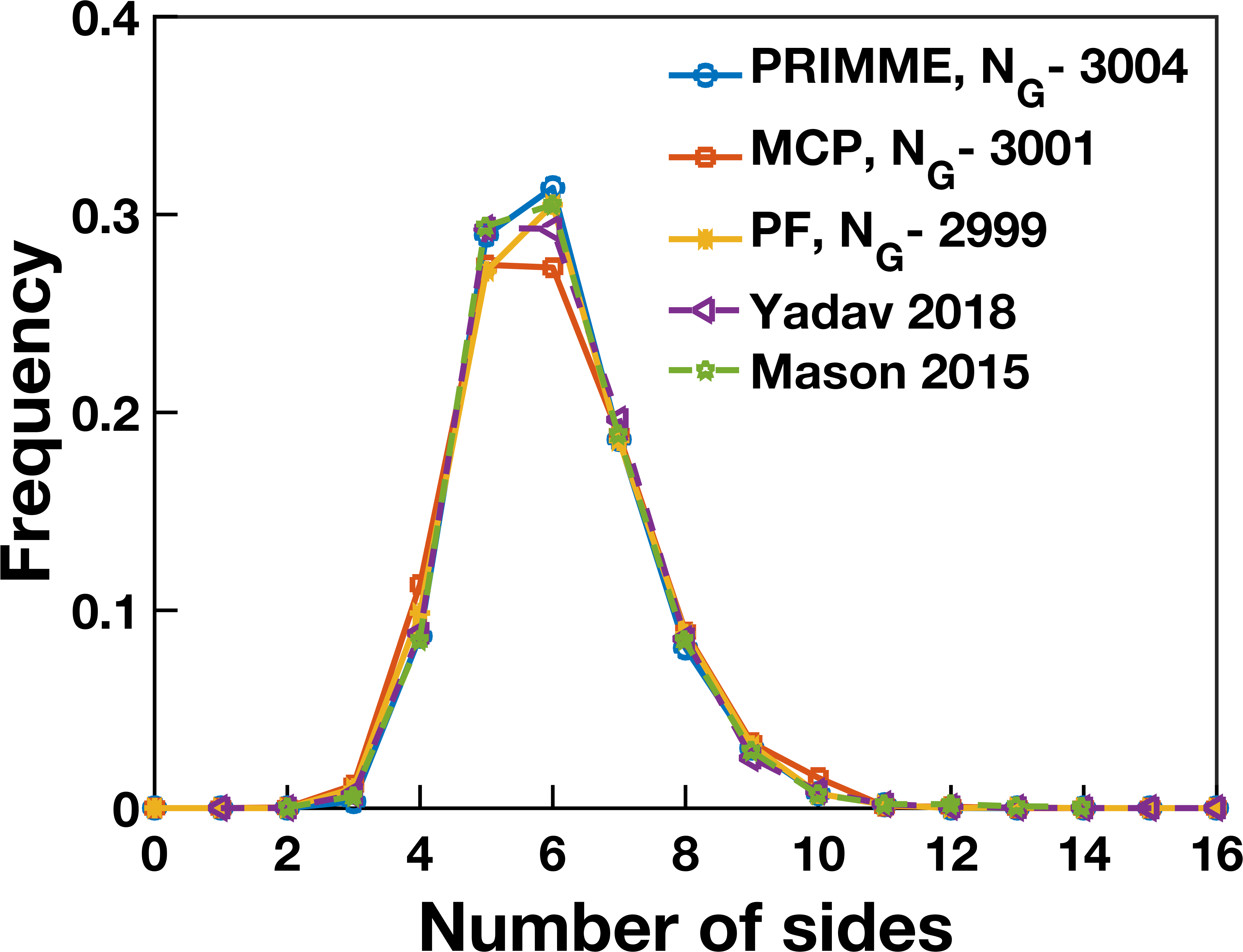} 	\includegraphics[width=0.32\textwidth]{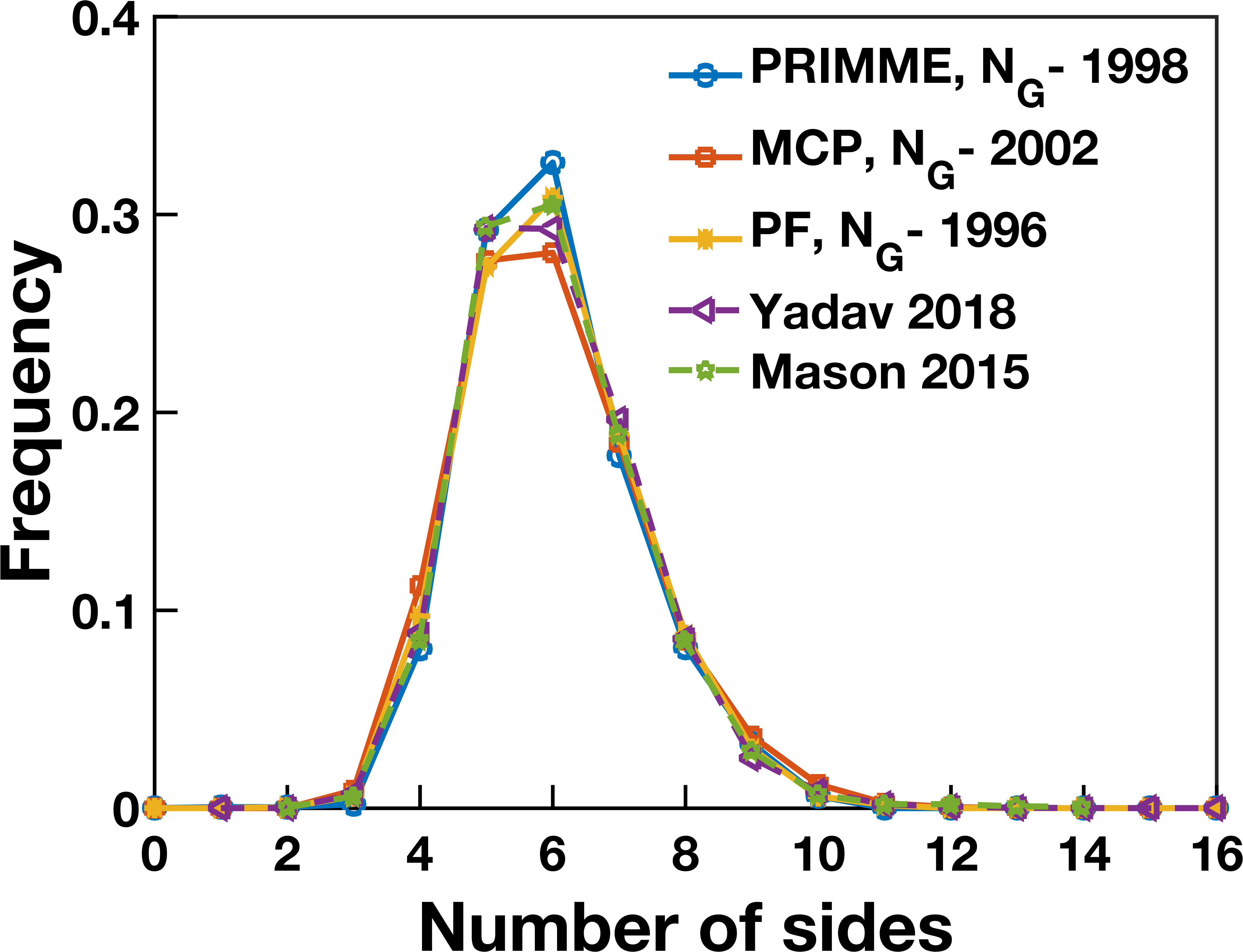} 	\caption{\label{fig:num_sides_distr}}
    \end{subfigure}
    \caption{Topological analysis of the $2400\times2400$ pixel domain with 20,000 initial grains. (a) Evolution of the mean number of sides with time from PRIMME, MCP, and PF. The PRIMME and MCP results used the scaling from the geometric analysis. (b) Comparison of the distribution of the number of sides for approximately 4000, 3000, and 2000 grains from PRIMME, MCP, and PF. The results from Yadav 2018 \cite{Yadav2018} and Mason 2015 \cite{Mason2015} are also included, for reference.}
\end{figure}
The mean number of sides $\langle F \rangle$ versus time from PRIMME, and MCP and PF is shown in Fig.~\ref{fig:av_num_sides_vs_time}. According to Euler's theorem, the average number of sides during steady-state growth should be six \cite{THOMPSON2001}.  PRIMME correctly predicts this behavior, and its results are very similar to those from MCP and PF. 

The distribution of the number of sides for grain structures with approximately 4000, 3000, and 2000 grains from PRIMME, MCP, and PF are shown in Fig.~\ref{fig:num_sides_distr}. Like the grain size distribution, the distribution of the number of sides is also self similar during normal grain growth. PRIMME accurately predicts an unchanging distribution of the number of sides even as the grain structure evolves, and it predicts distributions that are very similar to those from MCP and PF. The PRIMME distributions are {in good agreement with} those from front tracking simulations \cite{Mason2015} and PF simulations \cite{Yadav2018} from the literature.
  
\subsubsection{von Neumann-Mullins relationship}

\begin{figure}[b!]
\centering
 \includegraphics{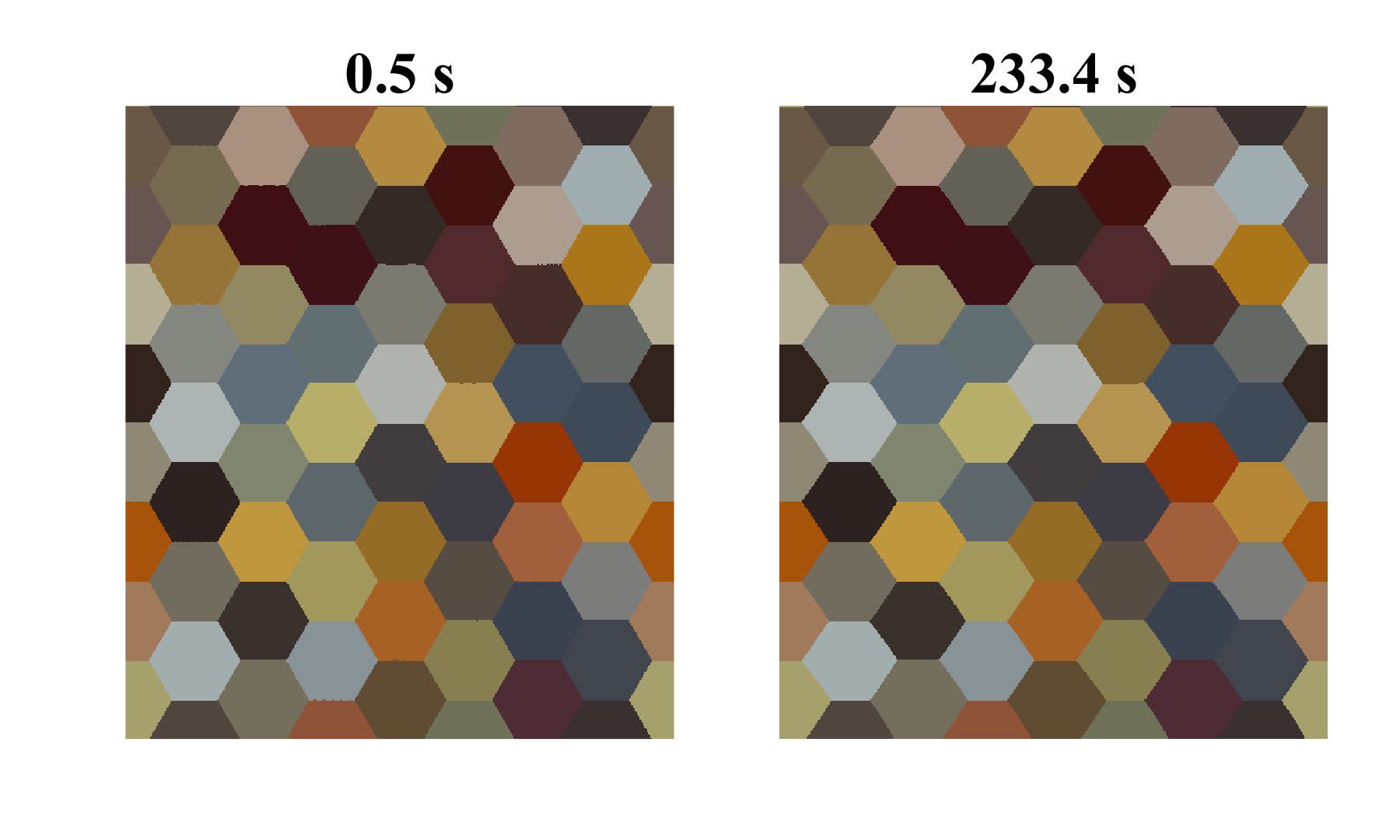}
 \caption{Testing von Neumann-Mullins relationship with hexagonal grain structure. PRIMME accurately simulates no evolution of a hexagonal grain structure. }
 \label{FIG:hexagon}
\end{figure}

Finally, to test the von Neumann-Mullins relation using PRIMME, we simulate the grain growth of a $443 \times 512$ domain with $64$~hexaganal grains using periodic boundary conditions. The structure after 1 step  and after $500$~steps are shown in Fig.~\ref{FIG:hexagon}. PRIMME accurately predicts no evolution of the hexagonal grain structure. Note that PRIMME was not trained on any structures with hexagonal grains, nor on any rectangular domains.

The systematic investigation shows that PRIMME simulation results match with analytical models. Also, steady-state growth characteristic of normal grain growth is observed for large-scale two-dimensional simulations. Thus, PRIMME model is ready to simulate isotropic two-dimensional grain growth accurately.

\section{Discussion}

We have demonstrated that the newly developed PRIMME model is capable of simulating  isotropic two-dimentional grain growth. In this section, we discuss the model interpretability, the effects of regularization, error due to extrapolation, and the impact of overfitting. We also show PRIMME's ability to learn irregular grain growth and perform simulations accordingly. 

\subsection{Interpretability}
\begin{figure}[t!]
\centering
 \includegraphics[width=3.33in]{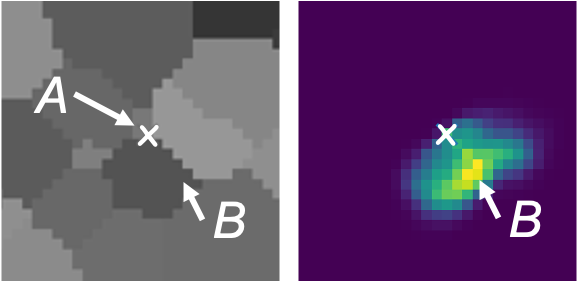}\\
 (a) \hspace{3cm} (b)
 \caption{Illustration of the interpretability of the PRIMME architecture. (a) shows the input grain structure and (b) shows the output action likelihood of the PRIMME neural network.Colors represent the action likelihood for site A to obtain the same grain number as that site after one step. The lightest color represents the largest action likelihood. In this example, site A flips to the site B grain number in the next step.}
 \label{FIG:interpret}
\end{figure}

A major benefit of PRIMME is that it is not just a black box. The architecture of PRIMME is built specifically to increase interpretability by predicting the action likelihood rather than the action itself. Figure~\ref{FIG:interpret} illustrates how we can interpret what microstructural features are critical to flip a site A from grain $i$ to grain $j$. The neural network calculates the action likelihood for each site to flip to the same  grain as other pixels. Figure~\ref{FIG:interpret}(b) shows the action likelihood map calculated for site A. In this example, site A will flip to the grain that contains site B because that is the action with the greatest likelihood, according to the neural network. Interestingly, site B is not a neighbor of site A but is near a neighboring triple junction, suggesting that PRIMME is using information from triple junctions to determine grain boundary motion. This type of information can be assessed for each site and evaluated to determine statistically what features are critical to grain growth. 

\subsection{The Effects of Regularization}

\begin{figure*}[p]
\centering
 \includegraphics[width=5.5in]{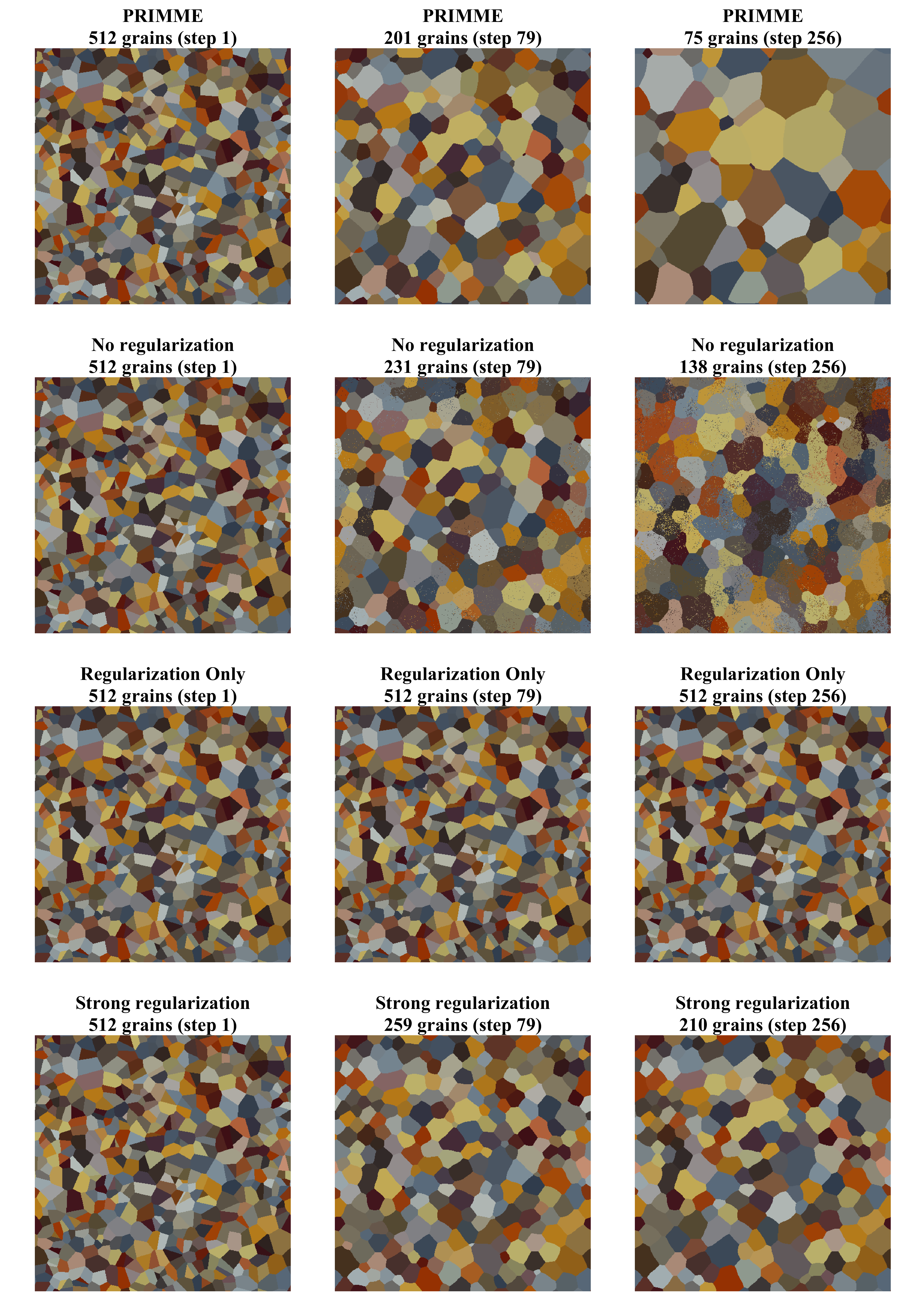}
 \caption{Effects of regularization on PRIMME microstructure growth in the $512\times512$ pixel domain with $512$~initial grains. The first row shows the standard PRIMME results from Fig.~\ref{FIG:growth}, for reference. The predicted behavior with no regularization, only regularization, and strong regularization are shown in the second, third, and fourth rows, respectively.}
 \label{FIG:regularization}
\end{figure*}

The physics-informed regularization is one of the key components that enables PRIMME to predict accurate grain growth. To illustrate its importance, we simulate the grain growth in the $512\times512$ pixel domain with 512 initial grains with three changes to  the regularization, as shown in  Fig.~\ref{FIG:regularization}, to demonstrate how it affects the predicted growth. Fig.~\ref{FIG:regularization} also contains the standard PRIMME results from Fig.~\ref{FIG:growth}, for reference. 

When regularization is removed (i.e.,  $\lambda=0$ in Eq.~\eqref{eq:loss}), sites in the middle of grains will occasionally flip to some other grain number, as shown in the second row of Fig.~\ref{FIG:regularization}.  As a result, the grains evolve into non-convex shapes and divide into discontiguous sections. Such behaviors are not consistent with normal isotropic grain growth. 

When only regularization is applied (i.e., effectively $\lambda=\infty$ in Eq.~\eqref{eq:loss}), nothing is learned from the training data. Under this condition, the microstructure fails to evolve significantly, as shown in the third row of Fig.~\ref{FIG:regularization}, since a single step does not reduce the regularization. The smallest grains evolve slightly but never successfully disappear. 

When the regularization loss is weighted approximately $32$~times more than the training loss (i.e.,  $\lambda=32$ in Eq.~\eqref{eq:loss}), the grain structure initially evolves, but the grain boundary migration slows and eventually stops, as shown in the last row of Fig.~\ref{FIG:regularization}. This demonstrates that the regularization and data-driven training are both equally necessary to achieve accurate grain growth behavior.  

\subsection{Error from Extrapolating Outside the Training Data Set}

\begin{figure*}[b!]

\centering
	\begin{subfigure}[b]{0.45\textwidth}
 		\includegraphics[width=\textwidth]{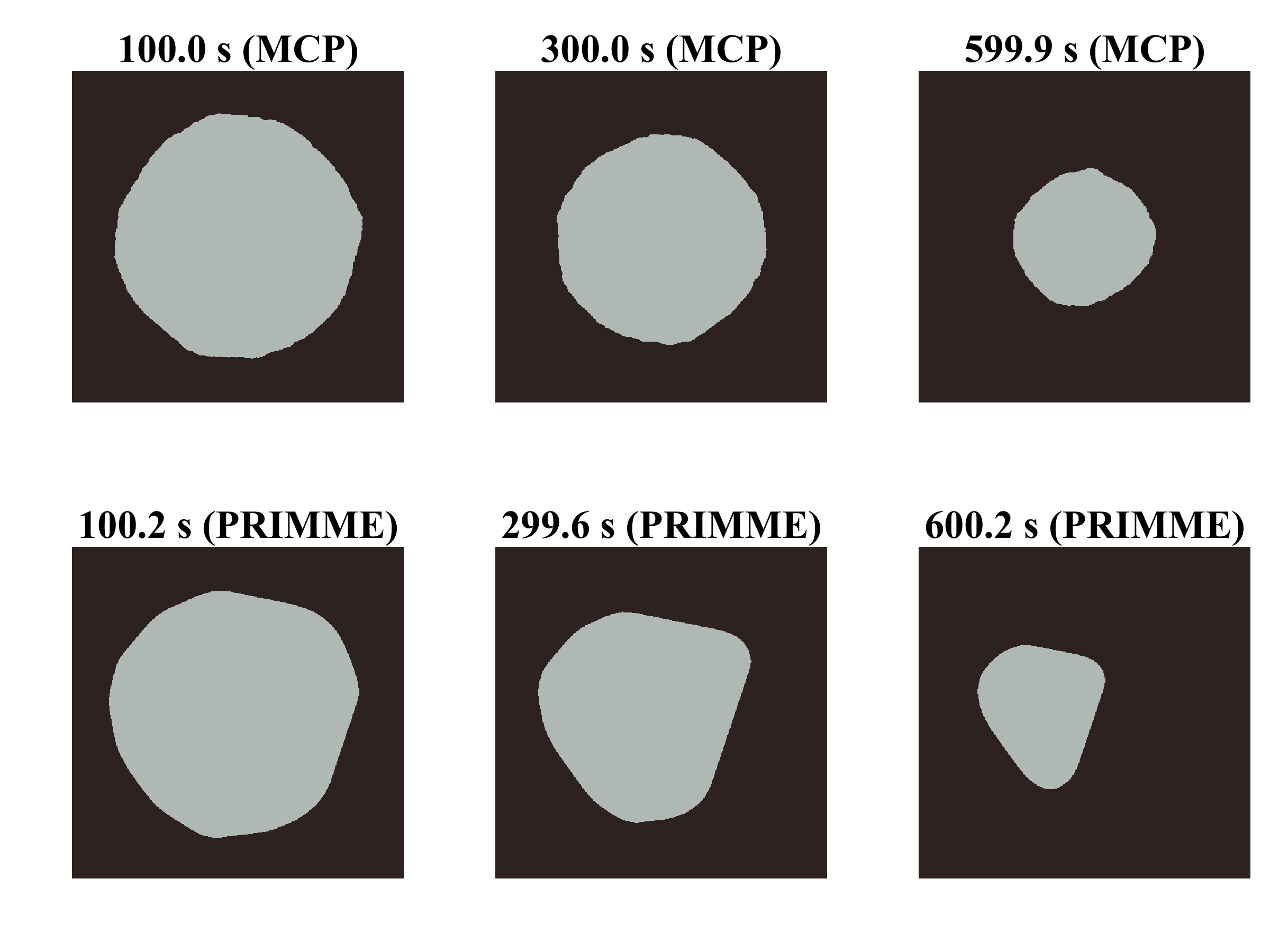} 		
	    \caption{\label{FIG:vanish200_images}}
	\end{subfigure}	\hspace{0.1in}
	\begin{subfigure}[b]{0.45\textwidth}
 		\includegraphics[width=\textwidth]{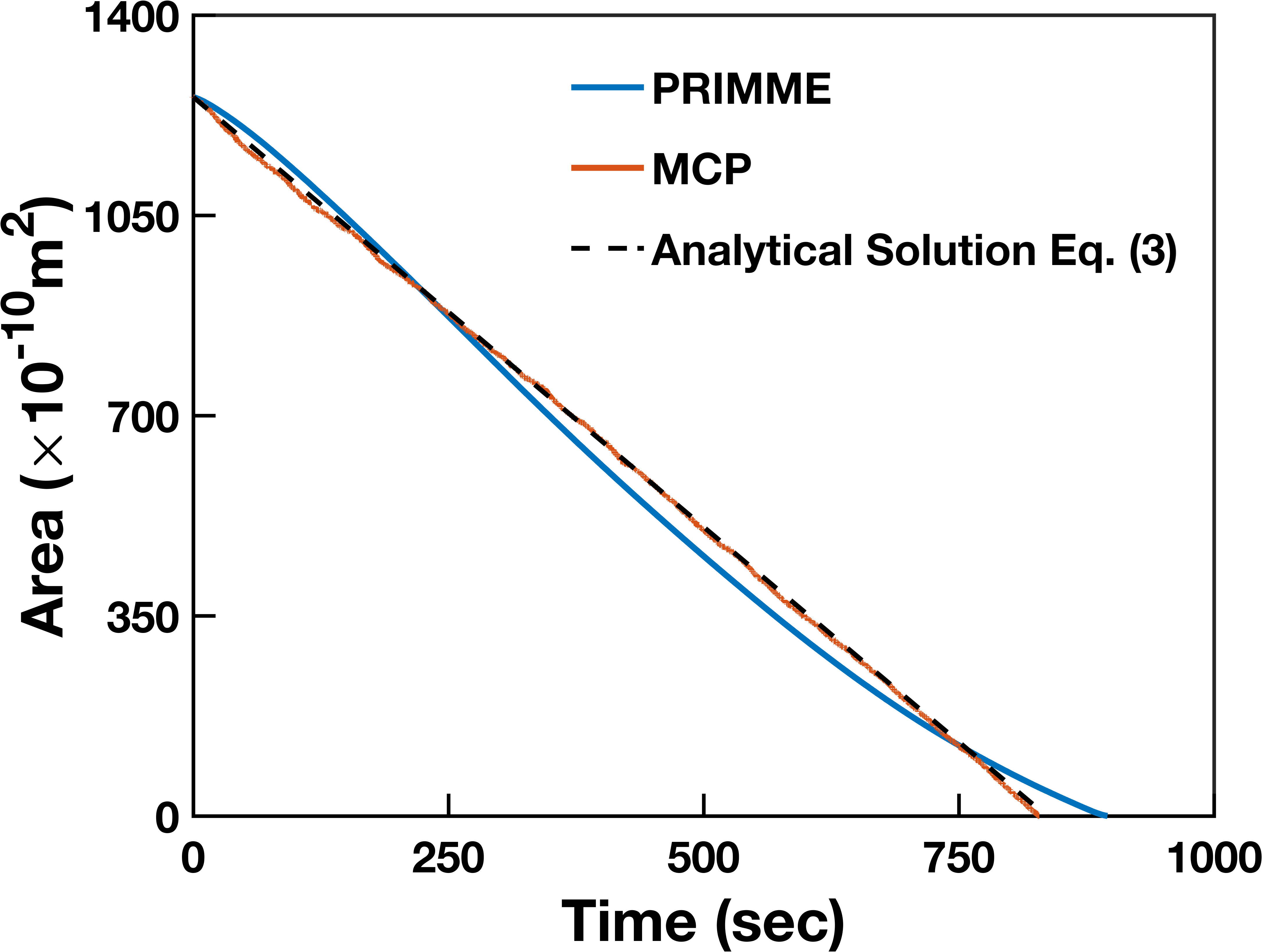} 		
	    \caption{\label{FIG:vanish200_plot}}
	\end{subfigure}

  \caption{Effects of extrapolation on PRIMME modeling the shrinking of a 200 pixel radius circular grain embedded in a $512\times512$ pixel matrix. (a) Images of the shrinking grain using MCP and PRIMME. (b) The area of the circular grain over time predicted by PRIMME and MCP, with the analytical solution from Eq.~\eqref{eq:A(t)} shown for reference.}
 \label{FIG:vanish200}
\end{figure*}

Machine learning models learn from training data. This enables PRIMME to learn grain growth behavior without a definition of the underlying physics. Yet, PRIMME is limited by the behavior it observes.  Due to computational constraints, PRIMME was trained with relatively small grain sizes, as shown in Fig.~\ref{FIG:training_size_distribution}. To understand what occurs as grain sizes increase, we apply PRIMME to simulate the behavior of a $R=200$ pixel radius circular grain embedded in $512\times 512$ pixel matrix. This circular grain is ten times larger in size than any of the grains in the training data. Its evolution is shown in Fig.~\ref{FIG:vanish200_images}. As the simulation progresses, the circular shape of the grain is not maintained as occurred with the smaller circular grain from Fig.~\ref{FIG:vanish}, but it develops more distinct faceted boundary segments. Also, the relationship between the its area and time is no longer linear (Fig.~\ref{FIG:vanish200_plot}). This result likely demonstrates error introduced into the prediction due to extrapolation to cases far outside of the training set.

\subsection{Overtraining and Extrapolation}

\begin{figure}[b!]
\centering
 \includegraphics[width=0.45\textwidth]{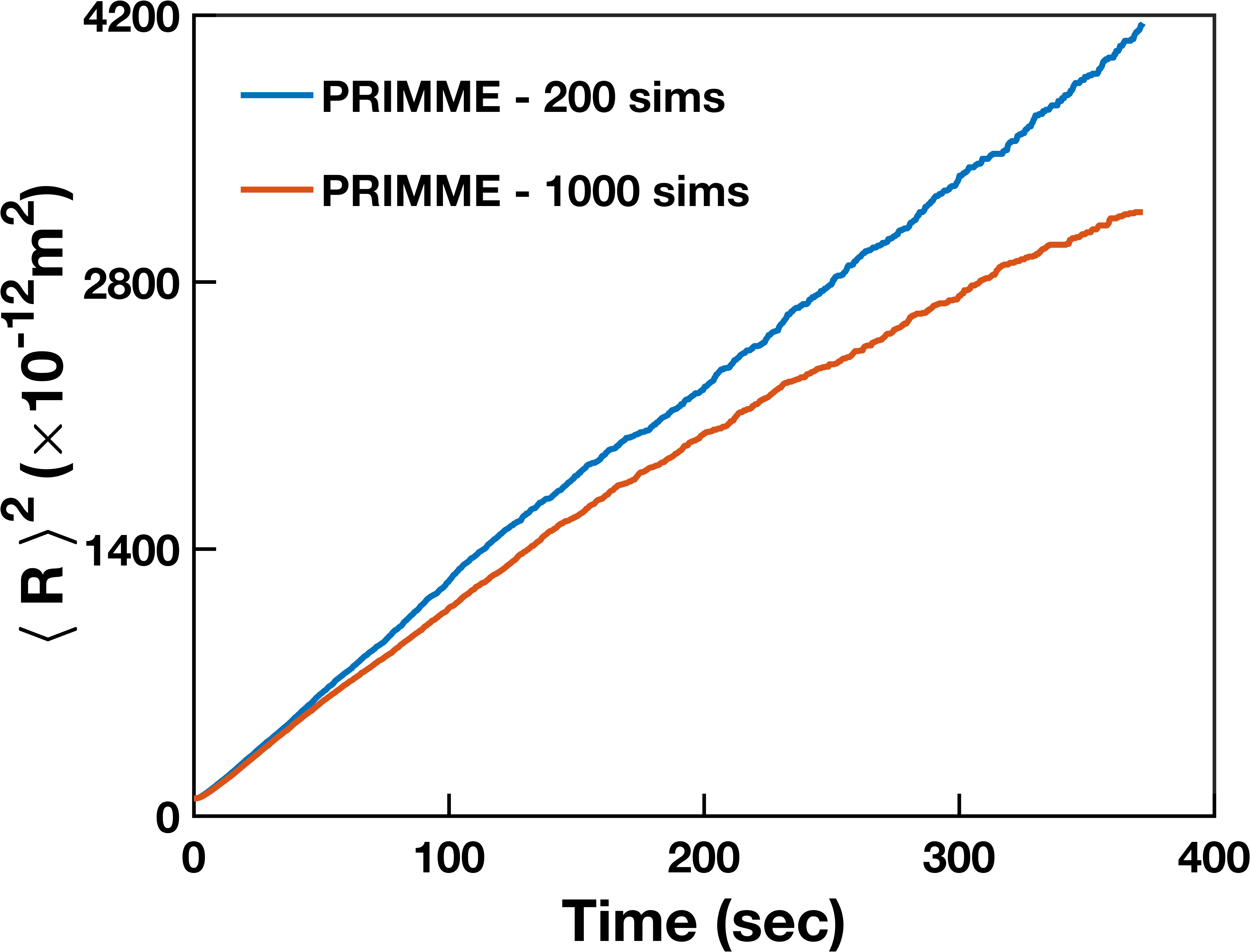}
 \caption{Effects of overfitting on the grain growth predicted by PRIMME for the $2400\times2400$ pixel domain with 20,000 initial grains. The square of the average grain size with time is shown for PRIMME trained with 200 simulations and PRIMME trained with 1000 simulations. }
 \label{FIG:overtraining}
\end{figure}

As the neural network is trained, it improves its performance when replicating behavior similar to its training data, i.e. interpolation. However, this occurs at the cost of a less effective prediction of behavior outside the training data, i.e.\ extrapolation. This is effectively caused by the neural network overfitting to the training data. Figure~\ref{FIG:overtraining} compares the grain growth behavior in a 20,000 grain polycrystal predicted by PRIMME trained with $200$~simulations to PRIMME trained with $1000$~simulations. PRIMME trained with 200 simulations predicts the correct behavior, while PRIMME trained with 1000 simulations predicts deviation from parabolic growth after 200 s. We hypothesize that the extrapolated behavior at longer times, once the grain size gets outside of the range in the training data, becomes more dependent on the regularization than the data as we overfit. Hence, large grains begin to stop growing, as observed in Fig.~\ref{FIG:regularization}.

\subsection{Modeling of Irregular Grain Growth}

A major benefit of PRIMME is that it is trained by data without governing assumptions about grain growth, suggesting it could directly learn from contiguous experimental data. As such, PRIMME has the potential to predict currently unexplained behavior by training with experimental datasets that exhibit irregular grain growth. We demonstrate how PRIMME can be trained to predict irregular grain growth by training using MCP model data similar to that discussed in Section \ref{ssec:sppark}, but using $kT=0$ rather than 0.5. 

It is well-established in the literature that using $kT=0$ in MCP simulations results in pinning of the grain boundaries to lattice sites \cite{zollner2014new}, which causes grain boundary faceting that is not representative of normal grain growth. While the predicted grain growth behavior with $kT=0$ is non-physical, it does provide a convenient source of irregular grain growth data. 

\begin{figure}[b!]
\centering
	\begin{subfigure}[b]{0.45\textwidth}
 		\includegraphics[width=\textwidth]{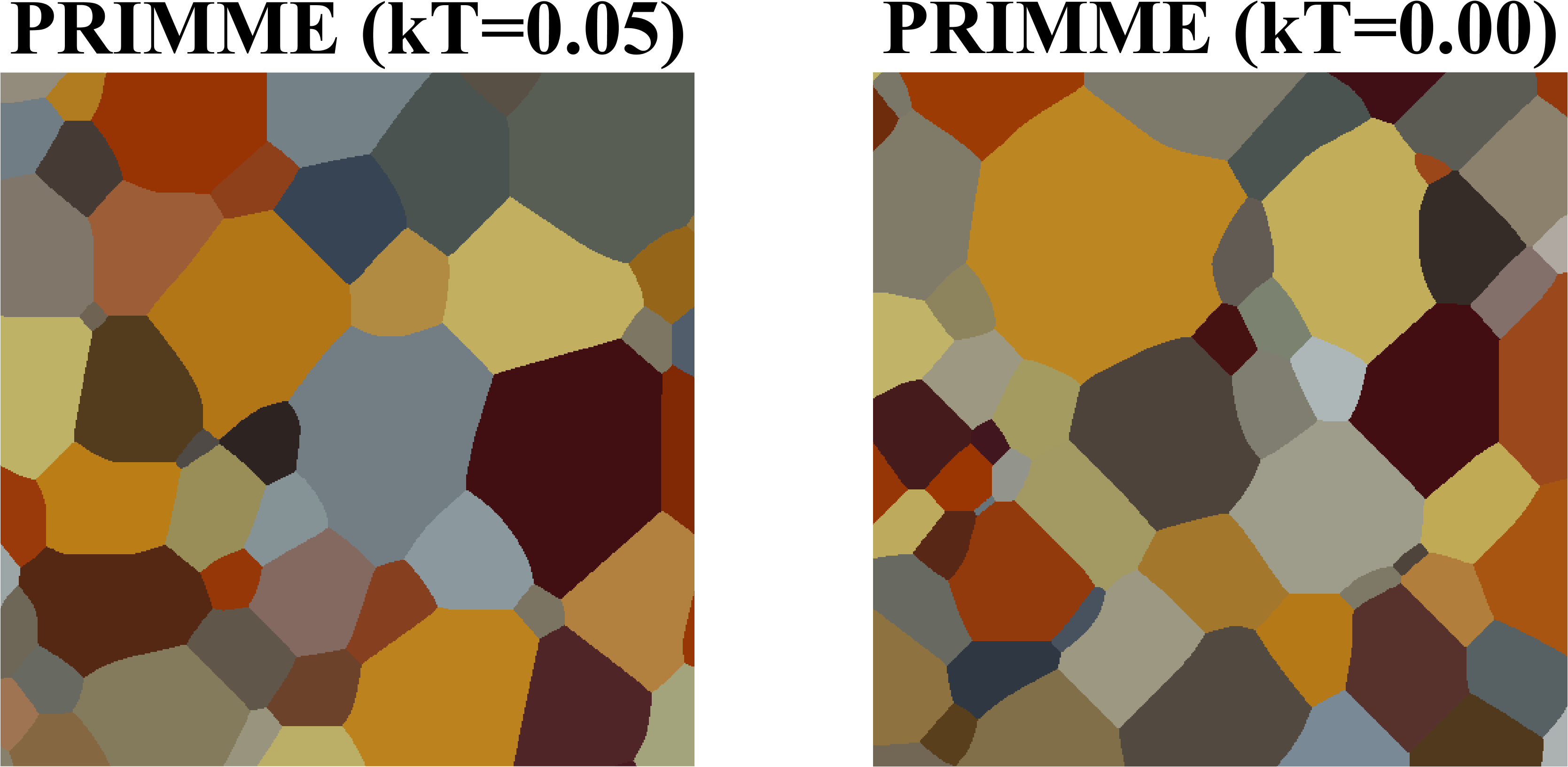} 		
	    \caption{\label{FIG:growthIr_images}}
	\end{subfigure}	
	
	\begin{subfigure}[b]{0.45\textwidth}
 		\includegraphics[width=\textwidth]{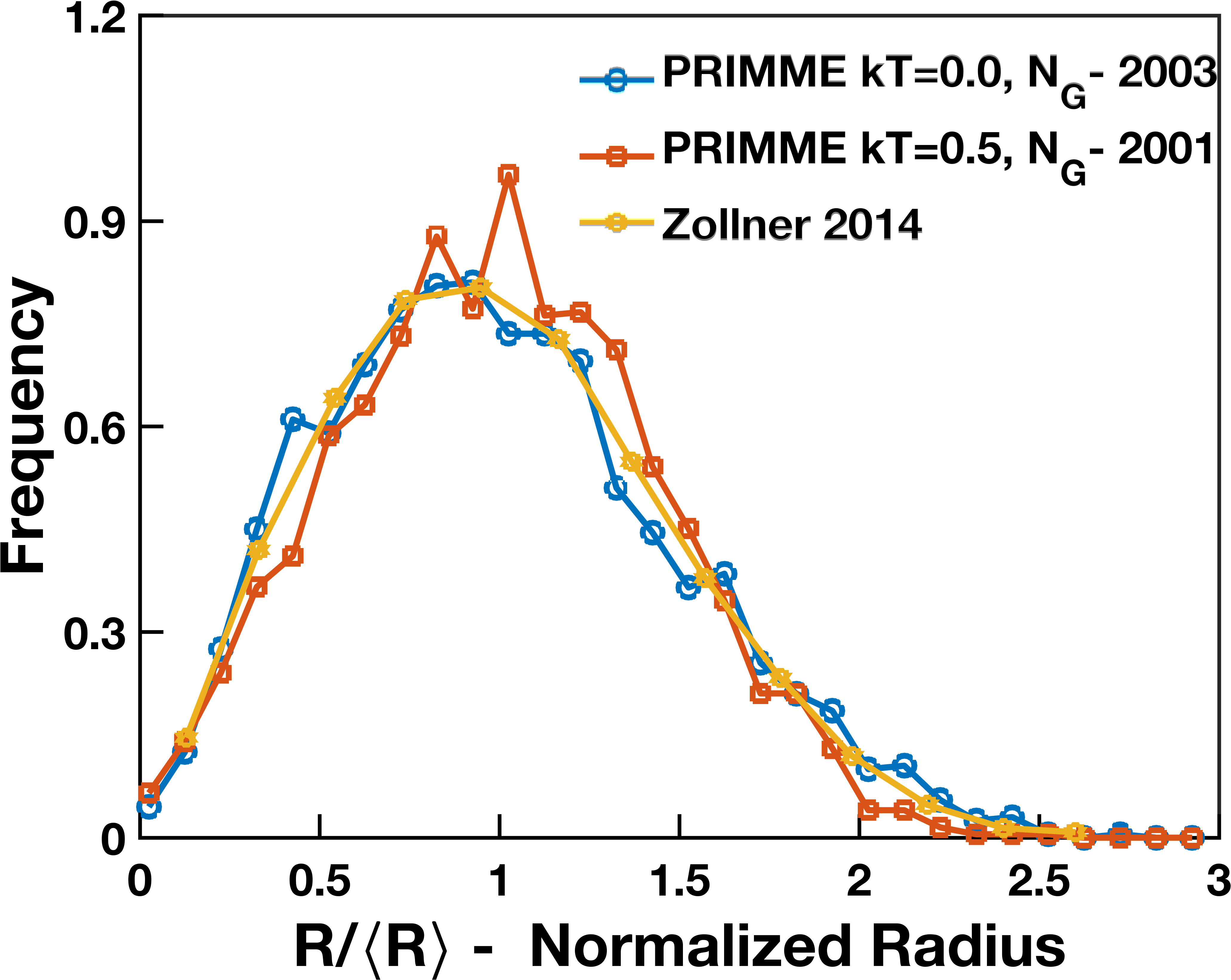} 		
	    \caption{\label{FIG:growthIr_plot}}
	\end{subfigure}
 
 \caption{Comparison of irregular grain growth with normal grain growth using results from the $2400\times2400$ pixel polycrystal with 20,000 initial grains. (a) Zoomed in $512\times512$ subdomain once the structure has evolved to 1000 grains from regular PRIMME (trained with $kT=0.5$ MCP data) compared with irregular PRIMME (trained with $kT=0$ MCP data).  (b) Comparison of the normalized grain size distribution once the structure has reached $N_G=2000$ grains. Regular PRIMME is compared with irregular PRIMME and with an MCP simulation from the literature with $kT=0$ \cite{zollner2014new}.}
 \label{FIG:growthIr}
\end{figure}

Figure~\ref{FIG:growthIr_images} shows a zoomed in view of PRIMME results for the $2400\times2400$ pixel polycrystal with 20,000 initial grains trained using the MCP model data with $kT=0$ that results in  irregular grain growth. The $kT=0$ PRIMME predicts more rectangular grains, with some triple junction angles near 90$^\circ$. It also predicts a few large grains and a cluster of small grains.

Figure~\ref{FIG:growthIr_plot} compares the normalized grain size distribution of irregular PRIMME with regular PRIME and MCP results with $kT=0.0$ from the literature.  The shape of the normalized grain size  distribution is asymmetrical for irregular PRIMME and its peak is shifted  towards small grains compared to regular grain growth. The grain size distribution of irregular PRIMME has good agreement with previous MCP simulation with $kT=0$ \cite{zollner2014new}, suggesting that just by changing the training data, PRIMME can predict distinct grain boundary migration

While the irregular grain growth behavior generated using the MCP model with $kT = 0$ is artificial, it does demonstrate the potential for PRIMME to be trained to predict irregular grain growth. It could be trained using experimental data or results from simulations that account for grain boundary anisotropy.

\section{Conclusions}
\label{sec:conclusions}

We have successfully developed the PRIMME model, a machine learning neural network with regularization to make site-wise action predictions for grain evolution. PRIMME, which was trained on simulated grain structures generated using the MCP model in SPPARKS,  is capable of simulating  isotropic two-dimensional grain growth {in any rectangular domain with any initial grain structure. Its predictions are in good agreement with analytical models and with the results from physics-based isotropic grain growth models from the literature. We demonstrated this agreement for grain structures with 512 initial grains in a $512\times512$ pixel domain and 20,000 initial grains in a $2400\times2400$ pixel domain. Predictions of a shrinking grain showed good agreement with respect to the change in area with time, but the shrinking grain took on an oblong shape that is not it agreement with other deterministic methods such as the phase field method}. Results produced by PRIMME are interpretable, using the action likelihood for each site{, and i}t can be taught to predict irregular grain growth using MCP training data with $kT=0$. {V}ariations in regularization and training data {significantly} affect the results{, and need to be investigated more in the future}. Computational resource requirement for large scale simulations is comparable or less than PF and MCP models. 

Our results are among the first demonstrations of a machine learning model replicating isotropic grain growth from training with physics-based simulation data. However, future work is necessary to address challenges with extrapolation without overfitting. PRIMME may also be more effectively trained using data from deterministic simulation methods, such as from phase field models. PRIMME could be extended  to reproduce realistic three-dimensional complex grain growth, where grain boundary energy and mobility are anisotropic, after training with experimental data. {Furthermore, PRIMME could be extended to model grain growth and densification during sintering, though new capability would need to be added to ensure conservation of mass.} Thus, our results demonstrate a huge potential for machine learning to be used to predict microstructure evolution for various materials and applications, which is currently very difficult to do with analytical or computational models.

\section*{Acknowledgements}

The authors would like to acknowledge financial support by the U.S. Department of Energy, Office of Science, Basic Energy Sciences under Award \#DE-SC0020384. This material is also based upon work supported by the U.S. Department of Defence through a Science, Mathematics, and Research for Transformation (SMART) scholarship. 

\bibliography{mybibfile}

\begin{thebibliography}{50}
\expandafter\ifx\csname natexlab\endcsname\relax\def\natexlab#1{#1}\fi
\providecommand{\url}[1]{\texttt{#1}}
\providecommand{\href}[2]{#2}
\providecommand{\path}[1]{#1}
\providecommand{\DOIprefix}{doi:}
\providecommand{\ArXivprefix}{arXiv:}
\providecommand{\URLprefix}{URL: }
\providecommand{\Pubmedprefix}{pmid:}
\providecommand{\doi}[1]{\href{http://dx.doi.org/#1}{\path{#1}}}
\providecommand{\Pubmed}[1]{\href{pmid:#1}{\path{#1}}}
\providecommand{\bibinfo}[2]{#2}
\ifx\xfnm\relax \def\xfnm[#1]{\unskip,\space#1}\fi
\bibitem[{Dillon et~al.(2016)Dillon, Tai, and Chen}]{dillon2016importance}
\bibinfo{author}{S.~J. Dillon}, \bibinfo{author}{K.~Tai},
  \bibinfo{author}{S.~Chen},
\newblock \bibinfo{title}{The importance of grain boundary complexions in
  affecting physical properties of polycrystals},
\newblock \bibinfo{journal}{Current Opinion in Solid State and Materials
  Science} \bibinfo{volume}{20} (\bibinfo{year}{2016})
  \bibinfo{pages}{324--335}.
\bibitem[{Humphreys(1997)}]{humphreys1997unified}
\bibinfo{author}{F.~Humphreys},
\newblock \bibinfo{title}{A unified theory of recovery, recrystallization and
  grain growth, based on the stability and growth of cellular
  microstructures—i. the basic model},
\newblock \bibinfo{journal}{Acta Materialia} \bibinfo{volume}{45}
  (\bibinfo{year}{1997}) \bibinfo{pages}{4231--4240}.
\bibitem[{Burke and Turnbull(1952)}]{Burke1952}
\bibinfo{author}{J.~Burke}, \bibinfo{author}{D.~Turnbull},
\newblock \bibinfo{title}{Recrystallization and grain growth},
\newblock \bibinfo{journal}{Progress in Metal Physics} \bibinfo{volume}{3}
  (\bibinfo{year}{1952}) \bibinfo{pages}{220--292}.
\bibitem[{Mullins(1956)}]{mullins1956two}
\bibinfo{author}{W.~W. Mullins},
\newblock \bibinfo{title}{Two-dimensional motion of idealized grain
  boundaries},
\newblock \bibinfo{journal}{Journal of Applied Physics} \bibinfo{volume}{27}
  (\bibinfo{year}{1956}) \bibinfo{pages}{900--904}.
\bibitem[{Wu(1982)}]{wu1982potts}
\bibinfo{author}{F.-Y. Wu},
\newblock \bibinfo{title}{The potts model},
\newblock \bibinfo{journal}{Reviews of modern physics} \bibinfo{volume}{54}
  (\bibinfo{year}{1982}) \bibinfo{pages}{235}.
\bibitem[{Anderson et~al.(1984)Anderson, Srolovitz, Grest, and
  Sahni}]{anderson1984computer}
\bibinfo{author}{M.~Anderson}, \bibinfo{author}{D.~Srolovitz},
  \bibinfo{author}{G.~Grest}, \bibinfo{author}{P.~Sahni},
\newblock \bibinfo{title}{Computer simulation of grain growth—i. kinetics},
\newblock \bibinfo{journal}{Acta metallurgica} \bibinfo{volume}{32}
  (\bibinfo{year}{1984}) \bibinfo{pages}{783--791}.
\bibitem[{Srolovitz et~al.(1984)Srolovitz, Anderson, Sahni, and
  Grest}]{srolovitz1984computer}
\bibinfo{author}{D.~Srolovitz}, \bibinfo{author}{M.~P. Anderson},
  \bibinfo{author}{P.~S. Sahni}, \bibinfo{author}{G.~S. Grest},
\newblock \bibinfo{title}{Computer simulation of grain growth—ii. grain size
  distribution, topology, and local dynamics},
\newblock \bibinfo{journal}{Acta metallurgica} \bibinfo{volume}{32}
  (\bibinfo{year}{1984}) \bibinfo{pages}{793--802}.
\bibitem[{Fan and Chen(1997)}]{fan1997computer}
\bibinfo{author}{D.~Fan}, \bibinfo{author}{L.-Q. Chen},
\newblock \bibinfo{title}{Computer simulation of grain growth using a continuum
  field model},
\newblock \bibinfo{journal}{Acta Materialia} \bibinfo{volume}{45}
  (\bibinfo{year}{1997}) \bibinfo{pages}{611--622}.
\bibitem[{Kim et~al.(2014)Kim, Kim, Dong, Steinbach, and Lee}]{kim2014phase}
\bibinfo{author}{H.-K. Kim}, \bibinfo{author}{S.~G. Kim},
  \bibinfo{author}{W.~Dong}, \bibinfo{author}{I.~Steinbach},
  \bibinfo{author}{B.-J. Lee},
\newblock \bibinfo{title}{Phase-field modeling for 3d grain growth based on a
  grain boundary energy database},
\newblock \bibinfo{journal}{Modelling and Simulation in Materials Science and
  Engineering} \bibinfo{volume}{22} (\bibinfo{year}{2014})
  \bibinfo{pages}{034004}.
\bibitem[{Miyoshi et~al.(2017)Miyoshi, Takaki, Ohno, Shibuta, Sakane,
  Shimokawabe, and Aoki}]{miyoshi2017ultra}
\bibinfo{author}{E.~Miyoshi}, \bibinfo{author}{T.~Takaki},
  \bibinfo{author}{M.~Ohno}, \bibinfo{author}{Y.~Shibuta},
  \bibinfo{author}{S.~Sakane}, \bibinfo{author}{T.~Shimokawabe},
  \bibinfo{author}{T.~Aoki},
\newblock \bibinfo{title}{Ultra-large-scale phase-field simulation study of
  ideal grain growth},
\newblock \bibinfo{journal}{NPJ Computational Materials} \bibinfo{volume}{3}
  (\bibinfo{year}{2017}) \bibinfo{pages}{1--6}.
\bibitem[{Miyoshi et~al.(2021)Miyoshi, Ohno, Shibuta, Yamanaka, and
  Takaki}]{Miyoshi2021M}
\bibinfo{author}{E.~Miyoshi}, \bibinfo{author}{M.~Ohno},
  \bibinfo{author}{Y.~Shibuta}, \bibinfo{author}{A.~Yamanaka},
  \bibinfo{author}{T.~Takaki},
\newblock \bibinfo{title}{Novel estimation method for anisotropic grain
  boundary properties based on bayesian data assimilation and phase-field
  simulation},
\newblock \bibinfo{journal}{Materials \& Design} \bibinfo{volume}{210}
  (\bibinfo{year}{2021}) \bibinfo{pages}{110089}.
\bibitem[{Chadwick and Voorhees(2021)}]{Chadwick2021}
\bibinfo{author}{A.~F. Chadwick}, \bibinfo{author}{P.~W. Voorhees},
\newblock \bibinfo{title}{The development of grain structure during additive
  manufacturing},
\newblock \bibinfo{journal}{Acta Materialia} \bibinfo{volume}{211}
  (\bibinfo{year}{2021}) \bibinfo{pages}{116862}.
\bibitem[{Moelans(2022)}]{Moelans2022}
\bibinfo{author}{N.~Moelans},
\newblock \bibinfo{title}{New phase-field model for polycrystalline systems
  with anisotropic grain boundary properties},
\newblock \bibinfo{journal}{Materials \& Design} \bibinfo{volume}{217}
  (\bibinfo{year}{2022}) \bibinfo{pages}{110592}.
\bibitem[{Liu et~al.(1996)Liu, Baudin, and Penelle}]{liu1996simulation}
\bibinfo{author}{Y.~Liu}, \bibinfo{author}{T.~Baudin},
  \bibinfo{author}{R.~Penelle},
\newblock \bibinfo{title}{Simulation of normal grain growth by cellular
  automata},
\newblock \bibinfo{journal}{Scripta Materialia} \bibinfo{volume}{34}
  (\bibinfo{year}{1996}).
\bibitem[{He et~al.(2006)He, Ding, Liu, and Shin}]{he2006computer}
\bibinfo{author}{Y.~He}, \bibinfo{author}{H.~Ding}, \bibinfo{author}{L.~Liu},
  \bibinfo{author}{K.~Shin},
\newblock \bibinfo{title}{Computer simulation of 2d grain growth using a
  cellular automata model based on the lowest energy principle},
\newblock \bibinfo{journal}{Materials Science and Engineering: A}
  \bibinfo{volume}{429} (\bibinfo{year}{2006}) \bibinfo{pages}{236--246}.
\bibitem[{Ding et~al.(2006)Ding, He, Liu, and Ding}]{ding2006cellular}
\bibinfo{author}{H.~Ding}, \bibinfo{author}{Y.~He}, \bibinfo{author}{L.~Liu},
  \bibinfo{author}{W.~Ding},
\newblock \bibinfo{title}{Cellular automata simulation of grain growth in three
  dimensions based on the lowest-energy principle},
\newblock \bibinfo{journal}{Journal of Crystal Growth} \bibinfo{volume}{293}
  (\bibinfo{year}{2006}) \bibinfo{pages}{489--497}.
\bibitem[{Xiong et~al.(2021)Xiong, Huang, Kafka, Lian, Yan, Chen, and
  Fang}]{Xiong2021}
\bibinfo{author}{F.~Xiong}, \bibinfo{author}{C.~Huang}, \bibinfo{author}{O.~L.
  Kafka}, \bibinfo{author}{Y.~Lian}, \bibinfo{author}{W.~Yan},
  \bibinfo{author}{M.~Chen}, \bibinfo{author}{D.~Fang},
\newblock \bibinfo{title}{Grain growth prediction in selective electron beam
  melting of ti-6al-4v with a cellular automaton method},
\newblock \bibinfo{journal}{Materials \& Design} \bibinfo{volume}{199}
  (\bibinfo{year}{2021}) \bibinfo{pages}{109410}.
\bibitem[{Baumard et~al.(2021)Baumard, Ayrault, Fandeur, Bordreuil, and
  Deschaux-Beaume}]{Baumard2021}
\bibinfo{author}{A.~Baumard}, \bibinfo{author}{D.~Ayrault},
  \bibinfo{author}{O.~Fandeur}, \bibinfo{author}{C.~Bordreuil},
  \bibinfo{author}{F.~Deschaux-Beaume},
\newblock \bibinfo{title}{Numerical prediction of grain structure formation
  during laser powder bed fusion of 316 l stainless steel},
\newblock \bibinfo{journal}{Materials \& Design} \bibinfo{volume}{199}
  (\bibinfo{year}{2021}) \bibinfo{pages}{109434}.
\bibitem[{Frost et~al.(1988)Frost, Thompson, Howe, and Whang}]{frost1988two}
\bibinfo{author}{H.~Frost}, \bibinfo{author}{C.~Thompson},
  \bibinfo{author}{C.~Howe}, \bibinfo{author}{J.~Whang},
\newblock \bibinfo{title}{A two-dimensional computer simulation of
  capillarity-driven grain growth: preliminary results},
\newblock \bibinfo{journal}{Scripta Metallurgica} \bibinfo{volume}{22}
  (\bibinfo{year}{1988}) \bibinfo{pages}{65--70}.
\bibitem[{Lazar et~al.(2010)Lazar, MacPherson, and Srolovitz}]{lazar2010more}
\bibinfo{author}{E.~A. Lazar}, \bibinfo{author}{R.~D. MacPherson},
  \bibinfo{author}{D.~J. Srolovitz},
\newblock \bibinfo{title}{A more accurate two-dimensional grain growth
  algorithm},
\newblock \bibinfo{journal}{Acta Materialia} \bibinfo{volume}{58}
  (\bibinfo{year}{2010}) \bibinfo{pages}{364--372}.
\bibitem[{Lazar et~al.(2011)Lazar, Mason, MacPherson, and
  Srolovitz}]{lazar2011more}
\bibinfo{author}{E.~A. Lazar}, \bibinfo{author}{J.~K. Mason},
  \bibinfo{author}{R.~D. MacPherson}, \bibinfo{author}{D.~J. Srolovitz},
\newblock \bibinfo{title}{A more accurate three-dimensional grain growth
  algorithm},
\newblock \bibinfo{journal}{Acta Materialia} \bibinfo{volume}{59}
  (\bibinfo{year}{2011}) \bibinfo{pages}{6837--6847}.
\bibitem[{Elsey et~al.(2009)Elsey, Esedog, Smereka et~al.}]{elsey2009diffusion}
\bibinfo{author}{M.~Elsey}, \bibinfo{author}{S.~Esedog},
  \bibinfo{author}{P.~Smereka}, et~al.,
\newblock \bibinfo{title}{Diffusion generated motion for grain growth in two
  and three dimensions},
\newblock \bibinfo{journal}{Journal of Computational Physics}
  \bibinfo{volume}{228} (\bibinfo{year}{2009}) \bibinfo{pages}{8015--8033}.
\bibitem[{Fausty et~al.(2021)Fausty, Murgas, Florez, Bozzolo, and
  Bernacki}]{Fausty2021}
\bibinfo{author}{J.~Fausty}, \bibinfo{author}{B.~Murgas},
  \bibinfo{author}{S.~Florez}, \bibinfo{author}{N.~Bozzolo},
  \bibinfo{author}{M.~Bernacki},
\newblock \bibinfo{title}{A new analytical test case for anisotropic grain
  growth problems},
\newblock \bibinfo{journal}{Applied Mathematical Modelling}
  \bibinfo{volume}{93} (\bibinfo{year}{2021}) \bibinfo{pages}{28--52}.
\bibitem[{Rollett et~al.(1989)Rollett, Srolovitz, and
  Anderson}]{rollett1989simulation}
\bibinfo{author}{A.~Rollett}, \bibinfo{author}{D.~J. Srolovitz},
  \bibinfo{author}{M.~Anderson},
\newblock \bibinfo{title}{Simulation and theory of abnormal grain
  growth—anisotropic grain boundary energies and mobilities},
\newblock \bibinfo{journal}{Acta metallurgica} \bibinfo{volume}{37}
  (\bibinfo{year}{1989}) \bibinfo{pages}{1227--1240}.
\bibitem[{McKenna et~al.(2014)McKenna, Poulsen, Lauridsen, Ludwig, and
  Voorhees}]{mckenna2014grain}
\bibinfo{author}{I.~McKenna}, \bibinfo{author}{S.~Poulsen},
  \bibinfo{author}{E.~M. Lauridsen}, \bibinfo{author}{W.~Ludwig},
  \bibinfo{author}{P.~W. Voorhees},
\newblock \bibinfo{title}{Grain growth in four dimensions: A comparison between
  simulation and experiment},
\newblock \bibinfo{journal}{Acta materialia} \bibinfo{volume}{78}
  (\bibinfo{year}{2014}) \bibinfo{pages}{125--134}.
\bibitem[{Bhattacharya et~al.(2021)Bhattacharya, Shen, Hefferan, Li, Lind,
  Suter, Krill~III, and Rohrer}]{bhattacharya2021grain}
\bibinfo{author}{A.~Bhattacharya}, \bibinfo{author}{Y.-F. Shen},
  \bibinfo{author}{C.~M. Hefferan}, \bibinfo{author}{S.~F. Li},
  \bibinfo{author}{J.~Lind}, \bibinfo{author}{R.~M. Suter},
  \bibinfo{author}{C.~E. Krill~III}, \bibinfo{author}{G.~S. Rohrer},
\newblock \bibinfo{title}{Grain boundary velocity and curvature are not
  correlated in ni polycrystals},
\newblock \bibinfo{journal}{Science} \bibinfo{volume}{374}
  (\bibinfo{year}{2021}) \bibinfo{pages}{189--193}.
\bibitem[{Carou et~al.(2022)Carou, Sartal, and Davim}]{Carou_undated-go}
\bibinfo{editor}{D.~Carou}, \bibinfo{editor}{A.~Sartal}, \bibinfo{editor}{J.~P.
  Davim} (Eds.), \bibinfo{title}{Machine Learning and Artificial Intelligence
  with Industrial Applications}, \bibinfo{publisher}{Springer International
  Publishing}, \bibinfo{address}{Gewerbestrasse}, \bibinfo{year}{2022}.
\bibitem[{Datta and Davim(2021)}]{Datta2021-un}
\bibinfo{editor}{S.~Datta}, \bibinfo{editor}{J.~P. Davim} (Eds.),
  \bibinfo{title}{Machine Learning in Industry}, Management and Industrial
  Engineering, \bibinfo{edition}{1} ed., \bibinfo{publisher}{Springer Nature},
  \bibinfo{address}{Cham, Switzerland}, \bibinfo{year}{2021}.
\bibitem[{Qian et~al.(2020)Qian, Kramer, Peherstorfer, and
  Willcox}]{qian2020lift}
\bibinfo{author}{E.~Qian}, \bibinfo{author}{B.~Kramer},
  \bibinfo{author}{B.~Peherstorfer}, \bibinfo{author}{K.~Willcox},
\newblock \bibinfo{title}{Lift \& learn: Physics-informed machine learning for
  large-scale nonlinear dynamical systems},
\newblock \bibinfo{journal}{Physica D: Nonlinear Phenomena}
  \bibinfo{volume}{406} (\bibinfo{year}{2020}) \bibinfo{pages}{132401}.
\bibitem[{Bostanabad et~al.(2018)Bostanabad, Zhang, Li, Kearney, Brinson,
  Apley, Liu, and Chen}]{bostanabad2018computational}
\bibinfo{author}{R.~Bostanabad}, \bibinfo{author}{Y.~Zhang},
  \bibinfo{author}{X.~Li}, \bibinfo{author}{T.~Kearney}, \bibinfo{author}{L.~C.
  Brinson}, \bibinfo{author}{D.~W. Apley}, \bibinfo{author}{W.~K. Liu},
  \bibinfo{author}{W.~Chen},
\newblock \bibinfo{title}{Computational microstructure characterization and
  reconstruction: Review of the state-of-the-art techniques},
\newblock \bibinfo{journal}{Progress in Materials Science} \bibinfo{volume}{95}
  (\bibinfo{year}{2018}) \bibinfo{pages}{1--41}.
\bibitem[{Chowdhury et~al.(2016)Chowdhury, Kautz, Yener, and
  Lewis}]{chowdhury2016image}
\bibinfo{author}{A.~Chowdhury}, \bibinfo{author}{E.~Kautz},
  \bibinfo{author}{B.~Yener}, \bibinfo{author}{D.~Lewis},
\newblock \bibinfo{title}{Image driven machine learning methods for
  microstructure recognition},
\newblock \bibinfo{journal}{Computational Materials Science}
  \bibinfo{volume}{123} (\bibinfo{year}{2016}) \bibinfo{pages}{176--187}.
\bibitem[{Bostanabad et~al.(2016)Bostanabad, Bui, Xie, Apley, and
  Chen}]{bostanabad2016stochastic}
\bibinfo{author}{R.~Bostanabad}, \bibinfo{author}{A.~T. Bui},
  \bibinfo{author}{W.~Xie}, \bibinfo{author}{D.~W. Apley},
  \bibinfo{author}{W.~Chen},
\newblock \bibinfo{title}{Stochastic microstructure characterization and
  reconstruction via supervised learning},
\newblock \bibinfo{journal}{Acta Materialia} \bibinfo{volume}{103}
  (\bibinfo{year}{2016}) \bibinfo{pages}{89--102}.
\bibitem[{de~Oca~Zapiain et~al.(2021)de~Oca~Zapiain, Stewart, and
  Dingreville}]{de2021accelerating}
\bibinfo{author}{D.~M. de~Oca~Zapiain}, \bibinfo{author}{J.~A. Stewart},
  \bibinfo{author}{R.~Dingreville},
\newblock \bibinfo{title}{Accelerating phase-field-based microstructure
  evolution predictions via surrogate models trained by machine learning
  methods},
\newblock \bibinfo{journal}{npj Computational Materials} \bibinfo{volume}{7}
  (\bibinfo{year}{2021}) \bibinfo{pages}{1--11}.
\bibitem[{Yang et~al.(2021)Yang, Cao, Zhang, Fan, Tang, Aberg, Sadigh, and
  Zhou}]{yang2021self}
\bibinfo{author}{K.~Yang}, \bibinfo{author}{Y.~Cao},
  \bibinfo{author}{Y.~Zhang}, \bibinfo{author}{S.~Fan},
  \bibinfo{author}{M.~Tang}, \bibinfo{author}{D.~Aberg},
  \bibinfo{author}{B.~Sadigh}, \bibinfo{author}{F.~Zhou},
\newblock \bibinfo{title}{Self-supervised learning and prediction of
  microstructure evolution with convolutional recurrent neural networks},
\newblock \bibinfo{journal}{Patterns} \bibinfo{volume}{2}
  (\bibinfo{year}{2021}) \bibinfo{pages}{100243}.
\bibitem[{James et~al.(2013)James, Witten, Hastie, and
  Tibshirani}]{james2013introduction}
\bibinfo{author}{G.~James}, \bibinfo{author}{D.~Witten},
  \bibinfo{author}{T.~Hastie}, \bibinfo{author}{R.~Tibshirani},
  \bibinfo{title}{An introduction to statistical learning}, volume
  \bibinfo{volume}{112}, \bibinfo{publisher}{Springer}, \bibinfo{year}{2013}.
\bibitem[{Mnih et~al.(2013)Mnih, Kavukcuoglu, Silver, Graves, Antonoglou,
  Wierstra, and Riedmiller}]{mnih2013playing}
\bibinfo{author}{V.~Mnih}, \bibinfo{author}{K.~Kavukcuoglu},
  \bibinfo{author}{D.~Silver}, \bibinfo{author}{A.~Graves},
  \bibinfo{author}{I.~Antonoglou}, \bibinfo{author}{D.~Wierstra},
  \bibinfo{author}{M.~Riedmiller},
\newblock \bibinfo{title}{Playing atari with deep reinforcement learning},
\newblock \bibinfo{journal}{arXiv preprint arXiv:1312.5602}
  (\bibinfo{year}{2013}).
\bibitem[{Cardona et~al.(2009)Cardona, Webb~III, Wagner, Tikare, Holm,
  Plimpton, Thompson, Slepoy, Zhou, Battaile et~al.}]{cardona2009crossing}
\bibinfo{author}{C.~G. Cardona}, \bibinfo{author}{E.~Webb~III},
  \bibinfo{author}{G.~Wagner}, \bibinfo{author}{V.~Tikare},
  \bibinfo{author}{E.~Holm}, \bibinfo{author}{S.~Plimpton},
  \bibinfo{author}{A.~Thompson}, \bibinfo{author}{A.~Slepoy},
  \bibinfo{author}{X.~Zhou}, \bibinfo{author}{C.~Battaile}, et~al.,
  \bibinfo{title}{Crossing the mesoscale no-man’s land via parallel kinetic
  Monte Carlo}, \bibinfo{type}{Technical Report}
  \bibinfo{number}{SAND2009-6226}, Sandia National Laboratories,
  \bibinfo{address}{Albuquerque, NM 87185}, \bibinfo{year}{2009}.
\bibitem[{Mnih et~al.(2015)Mnih, Kavukcuoglu, Silver, Rusu, Veness, Bellemare,
  Graves, Riedmiller, Fidjeland, Ostrovski et~al.}]{mnih2015human}
\bibinfo{author}{V.~Mnih}, \bibinfo{author}{K.~Kavukcuoglu},
  \bibinfo{author}{D.~Silver}, \bibinfo{author}{A.~A. Rusu},
  \bibinfo{author}{J.~Veness}, \bibinfo{author}{M.~G. Bellemare},
  \bibinfo{author}{A.~Graves}, \bibinfo{author}{M.~Riedmiller},
  \bibinfo{author}{A.~K. Fidjeland}, \bibinfo{author}{G.~Ostrovski}, et~al.,
\newblock \bibinfo{title}{Human-level control through deep reinforcement
  learning},
\newblock \bibinfo{journal}{nature} \bibinfo{volume}{518}
  (\bibinfo{year}{2015}) \bibinfo{pages}{529--533}.
\bibitem[{Hinton et~al.(2012)Hinton, Srivastava, Krizhevsky, Sutskever, and
  Salakhutdinov}]{hinton2012improving}
\bibinfo{author}{G.~E. Hinton}, \bibinfo{author}{N.~Srivastava},
  \bibinfo{author}{A.~Krizhevsky}, \bibinfo{author}{I.~Sutskever},
  \bibinfo{author}{R.~R. Salakhutdinov},
\newblock \bibinfo{title}{Improving neural networks by preventing co-adaptation
  of feature detectors},
\newblock \bibinfo{journal}{arXiv preprint arXiv:1207.0580}
  (\bibinfo{year}{2012}).
\bibitem[{Barmak et~al.(2013)Barmak, Eggeling, Kinderlehrer, Sharp, Ta’asan,
  Rollett, and Coffey}]{Barmak2013}
\bibinfo{author}{K.~Barmak}, \bibinfo{author}{E.~Eggeling},
  \bibinfo{author}{D.~Kinderlehrer}, \bibinfo{author}{R.~Sharp},
  \bibinfo{author}{S.~Ta’asan}, \bibinfo{author}{A.~Rollett},
  \bibinfo{author}{K.~Coffey},
\newblock \bibinfo{title}{Grain growth and the puzzle of its stagnation in thin
  films: The curious tale of a tail and an ear},
\newblock \bibinfo{journal}{Progress in Materials Science} \bibinfo{volume}{58}
  (\bibinfo{year}{2013}) \bibinfo{pages}{987--1055}.
\bibitem[{Permann et~al.(2020)Permann, Gaston, Andr{\v{s}}, Carlsen, Kong,
  Lindsay, Miller, Peterson, Slaughter, Stogner, and
  Martineau}]{permann2020moose}
\bibinfo{author}{C.~J. Permann}, \bibinfo{author}{D.~R. Gaston},
  \bibinfo{author}{D.~Andr{\v{s}}}, \bibinfo{author}{R.~W. Carlsen},
  \bibinfo{author}{F.~Kong}, \bibinfo{author}{A.~D. Lindsay},
  \bibinfo{author}{J.~M. Miller}, \bibinfo{author}{J.~W. Peterson},
  \bibinfo{author}{A.~E. Slaughter}, \bibinfo{author}{R.~H. Stogner},
  \bibinfo{author}{R.~C. Martineau},
\newblock \bibinfo{title}{{MOOSE}: Enabling massively parallel multiphysics
  simulation},
\newblock \bibinfo{journal}{{SoftwareX}} \bibinfo{volume}{11}
  (\bibinfo{year}{2020}) \bibinfo{pages}{100430}.
\bibitem[{Permann et~al.(2016)Permann, Tonks, Fromm, and
  Gaston}]{permann2016order}
\bibinfo{author}{C.~J. Permann}, \bibinfo{author}{M.~R. Tonks},
  \bibinfo{author}{B.~Fromm}, \bibinfo{author}{D.~R. Gaston},
\newblock \bibinfo{title}{Order parameter re-mapping algorithm for 3d phase
  field model of grain growth using fem},
\newblock \bibinfo{journal}{Computational Materials Science}
  \bibinfo{volume}{115} (\bibinfo{year}{2016}) \bibinfo{pages}{18--25}.
\bibitem[{Moelans et~al.(2008)Moelans, Blanpain, and
  Wollants}]{moelans2008quantitative}
\bibinfo{author}{N.~Moelans}, \bibinfo{author}{B.~Blanpain},
  \bibinfo{author}{P.~Wollants},
\newblock \bibinfo{title}{Quantitative analysis of grain boundary properties in
  a generalized phase field model for grain growth in anisotropic systems},
\newblock \bibinfo{journal}{Physical Review B} \bibinfo{volume}{78}
  (\bibinfo{year}{2008}) \bibinfo{pages}{024113}.
\bibitem[{Tikare et~al.(1998)Tikare, Holm, Fan, and
  Chen}]{tikare1998comparison}
\bibinfo{author}{V.~Tikare}, \bibinfo{author}{E.~Holm},
  \bibinfo{author}{D.~Fan}, \bibinfo{author}{L.-Q. Chen},
\newblock \bibinfo{title}{Comparison of phase-field and potts models for
  coarsening processes},
\newblock \bibinfo{journal}{Acta materialia} \bibinfo{volume}{47}
  (\bibinfo{year}{1998}) \bibinfo{pages}{363--371}.
\bibitem[{Suwa and Saito(2005)}]{suwa2005computer}
\bibinfo{author}{Y.~Suwa}, \bibinfo{author}{Y.~Saito},
\newblock \bibinfo{title}{Computer simulation of grain growth in three
  dimensions by the phase field model and the monte carlo method},
\newblock \bibinfo{journal}{Materials transactions} \bibinfo{volume}{46}
  (\bibinfo{year}{2005}) \bibinfo{pages}{1214--1220}.
\bibitem[{Yadav and Moelans(2018)}]{Yadav2018}
\bibinfo{author}{V.~Yadav}, \bibinfo{author}{N.~Moelans},
\newblock \bibinfo{title}{Comparison of coarsening behaviour in non-conserved
  and volume-conserved isotropic two-phase grain structures},
\newblock \bibinfo{journal}{Scripta Materialia} \bibinfo{volume}{146}
  (\bibinfo{year}{2018}) \bibinfo{pages}{142--145}.
\bibitem[{Z{\"o}llner(2016)}]{Zollner2016}
\bibinfo{author}{D.~Z{\"o}llner},
\newblock \bibinfo{title}{Grain microstructural evolution in 2d and 3d
  polycrystals under triple junction energy and mobility control},
\newblock \bibinfo{journal}{Computational Materials Science}
  \bibinfo{volume}{118} (\bibinfo{year}{2016}) \bibinfo{pages}{325--337}.
\bibitem[{Mason et~al.(2015)Mason, Lazar, MacPherson, and
  Srolovitz}]{Mason2015}
\bibinfo{author}{J.~K. Mason}, \bibinfo{author}{E.~A. Lazar},
  \bibinfo{author}{R.~D. MacPherson}, \bibinfo{author}{D.~J. Srolovitz},
\newblock \bibinfo{title}{Geometric and topological properties of the canonical
  grain-growth microstructure},
\newblock \bibinfo{journal}{Phys. Rev. E} \bibinfo{volume}{92}
  (\bibinfo{year}{2015}) \bibinfo{pages}{063308}.
\bibitem[{Thompson(2001)}]{THOMPSON2001}
\bibinfo{author}{C.~V. Thompson},
\newblock \bibinfo{title}{Grain growth and evolution of other cellular
  structures},
\newblock volume~\bibinfo{volume}{55} of \textit{\bibinfo{series}{Solid State
  Physics}}, \bibinfo{publisher}{Academic Press}, \bibinfo{year}{2001}, pp.
  \bibinfo{pages}{269--314}.
\bibitem[{Z{\"o}llner(2014)}]{zollner2014new}
\bibinfo{author}{D.~Z{\"o}llner},
\newblock \bibinfo{title}{A new point of view to determine the simulation
  temperature for the potts model simulation of grain growth},
\newblock \bibinfo{journal}{Computational materials science}
  \bibinfo{volume}{86} (\bibinfo{year}{2014}) \bibinfo{pages}{99--107}.

\end{thebibliography}











\end{document}